\documentclass[journal]{IEEEtran}
\usepackage{amsthm}
\usepackage{amsfonts}
\usepackage{amssymb}
\ifCLASSINFOpdf
  \usepackage[pdftex]{graphicx}
  \usepackage{graphicx}
  \graphicspath{{../pdf/}{../jpeg/}{../png/}}
  \DeclareGraphicsExtensions{.pdf,.jpeg,.png}
\else
  \usepackage[dvips]{graphicx}
  \graphicspath{{../eps/}}
  \DeclareGraphicsExtensions{.eps}
\fi
\ifCLASSOPTIONcompsoc
  \usepackage[caption=false,font=normalsize,labelfont=sf,textfont=sf]{subfig}
\else
\usepackage[caption=false,font=footnotesize]{subfig}
\fi
\usepackage{multirow}
\usepackage{epstopdf}
\usepackage{xcolor}
\usepackage{float}

\newtheorem{lem}{Lemma}
\usepackage{cite}
\ifCLASSINFOpdf
   \usepackage[pdftex]{graphicx}
   \else
\usepackage[dvips]{graphicx}
\fi
\usepackage[cmex10]{amsmath}
\usepackage[T1]{fontenc}
\newcommand*{\affmark}[1][*]{\textsuperscript{\dag}}
\usepackage{algorithm}
\usepackage{algpseudocode}
\usepackage{subfig}
\makeatletter
\def\BState{\State\hskip-\ALG@thistlm}
\makeatother
\begin{document}
\title{Robust Hybrid Transceiver Designs for Linear Decentralized Estimation in mmWave MIMO IoT Networks in the Face of Imperfect CSI}
\author{Priyanka Maity,~\IEEEmembership{Graduate Student Member,~IEEE,} Kunwar Pritiraj Rajput,~\IEEEmembership{Member,~IEEE,} Suraj Srivastava,~\IEEEmembership{Member,~IEEE,} Naveen K. D.  Venkategowda,~\IEEEmembership{ Member,~IEEE,}\\ Aditya K. Jagannatham,~\IEEEmembership{Senior Member,~IEEE} and Lajos Hanzo,~\IEEEmembership{Life Fellow,~IEEE}
 \thanks{L. Hanzo would like to acknowledge the financial support of the Engineering and Physical Sciences Research Council projects EP/W016605/1 and EP/X01228X/1 as well as of the European Research Council's Advanced Fellow Grant QuantCom (Grant No. 789028). The work of Aditya K. Jagannatham was supported in part by the
Qualcomm Innovation Fellowship, and in part by the Arun Kumar Chair.} \thanks{P. Maity, S. Srivastava and A. K. Jagannatham are with the
Department of Electrical Engineering, Indian Institute of Technology,
Kanpur, Kanpur, 208016, India (e-mail: pmaity@iitk.ac.in, ssrivast@iitk.ac.in, 
adityaj@iitk.ac.in.)}%

\thanks{K. P. Rajput was with  the Department of Electrical Engineering, Indian Institute of Technology, Kanpur, Kanpur, 208016, India. He is now with the interdisciplinary center of security reliability and trust (SnT), University of Luxembourg, 4365, Luxembourg (e-mail: kunwar.rajput@uni.lu)}%

\thanks{ N. K. D. Venkategowda is with the Department of Science and
Technology, Linköping University,  60174 Norrköping, Sweden (e-mail:
naveen.venkategowda@liu.se.)}%

\thanks{L. Hanzo is with the School of Electronics and Computer
Science, University of Southampton, Southampton SO17 1BJ, U.K.
(e-mail: lh@ecs.soton.ac.uk)}  }      

\maketitle
\begin{abstract}
Hybrid transceivers are designed for linear decentralized estimation
(LDE) in a mmWave multiple-input multiple-output (MIMO) IoT network
(IoTNe). For a noiseless fusion center (FC), it is demonstrated that
the MSE performance is determined by the number of RF chains used at
each IoT node (IoTNo). Next, the minimum-MSE RF transmit precoders
(TPCs) and receiver combiner (RC) matrices are designed for this setup
using the dominant array response vectors, and subsequently, a
closed-form expression is obtained for the baseband (BB) TPC at each IoTNo using
Cauchy's interlacing theorem. For a realistic noisy FC, it is shown
that the resultant mean squared error (MSE) minimization problem is
non-convex. To address this challenge, a block-coordinate
descent-based iterative scheme is proposed to obtain the fully digital
TPC and RC matrices followed by the simultaneous orthogonal matching
pursuit (SOMP) technique for decomposing the fully-digital transceiver
into its corresponding RF and BB components. A theoretical proof of
the convergence is also presented for the proposed iterative design
procedure. Furthermore, robust hybrid transceiver designs are also
derived for a practical scenario in the face of channel state
information (CSI) uncertainty. The centralized MMSE lower bound has
also been derived that benchmarks the performance of the proposed LDE
schemes. Finally, our numerical results characterize the performance
of the proposed transceivers as well as corroborate our various
analytical propositions.
\end{abstract}
%\vspace{-5pt}

\IEEEpeerreviewmaketitle
\vspace{-10pt}

\section{Introduction}
Internet of Things networks (IoTNes) constitute a key component of the Internet of Things (IoT) \cite{9071994}, which has the potential of supporting cutting-edge applications spanning diverse areas such as precision agriculture \cite{8521668}, remote healthcare \cite{9019876}, environmental monitoring \cite{8786991}, and smart cities \cite{9063400} etc.. IoTNes are comprised of miniature IoT nodes (IoTNos) dispersed over a wide geographical area for the transmission of their suitably pre-processed observations to a fusion centre (FC) over a wireless channel for further processing and estimation. The ever-increasing mobile device density coupled with the large number of data-intensive applications has led to the impending spectrum crunch in the sub 6 GHz band. This is further aggrevated by the deployment of a large number of IoT systems, each comprised of a multitude of IoTNos. \par
The abundant spectrum in the mmWave (30-300 GHz) band, which has hitherto only been partially explored, has the potential of supporting high data rates coupled with massive connectivity in the IoTNes. However, communication in the mmWave regime encounters several obstacles such as severe signal blockage and debilitating propagation losses  \cite{heath2016overview}. Fortunately, the low wavelength of mmWave signals renders it possible to integrate a large number of antennas into the otherwise small wireless devices. This in turn facilitates beamforming, which is capable of overcoming the aforementioned challenges in mmWave communication. The higher cost and power consumption due to the large number of RF chains required to support the increased number of antennas in a conventional architecture can be circumvented by employing innovative hybrid transceiver architectures \cite{molisch2017hybrid,el2014spatially,7397861,8332507}. Such a transceiver jointly performs analog TPC/ RC in the RF domain, together with digital TPC/ RC in the baseband, which minimizes the number of RF chains required. \par 
%\begin{table*}[t]
%\centering
%\caption{Boldly and explicitly contrasting our contributions in the literature}\label{Literature}
%\begin{tabular}{||l|c|c|c|c|c|c|c|c|c|c|c|c|c|c|c|c|}
%\hline
%Feature & [6] & [11] & [12] & [19],[20] & [22] & [23] & [26] &[27] &[28]&[31]&[32]&[38]& $\textbf{Our work}$ \\
%\hline
%\hline
%mmWave WSN &  &  &  &  &  &  & \checkmark  & \checkmark  &\checkmark & & & \checkmark &\checkmark\\
%\hline
%Vector parameter estimation & & \checkmark & \checkmark &  \checkmark  & \checkmark  & \checkmark & \checkmark & \checkmark && &\checkmark & &\checkmark\\
%\hline
%Per sensor power constraint &&\checkmark & &\checkmark  &  &  & \checkmark & \checkmark & \checkmark & & & \checkmark &\checkmark\\
%\hline
%Coherent MAC &  & \checkmark & \checkmark &  \checkmark &\checkmark &\checkmark &  &\checkmark & \checkmark & \checkmark& \checkmark & \checkmark &\checkmark\\
%\hline
%Hybrid transceiver design &\checkmark & & &  &  & & \checkmark & \checkmark &  & & & \checkmark &\checkmark\\
%\hline
%Noiseless FC  & & &  & &  & &   & & & & &  &\checkmark\\
%\hline
%Gaussian CSI uncertainty  & & &  & & & \checkmark &   & &  & \checkmark & \checkmark & \checkmark &\checkmark\\
%\hline
%\end{tabular}
%\end{table*} 
\begin{table*}[t]
\centering
\caption{Boldly and explicitly contrasting our contributions in the literature}\label{Literature}
\begin{tabular}{||l|c|c|c|c|c|c|c|c|c|c|c|c|c|c|c|c|}
\hline
Feature & [6] & [13] & [14] & [21],[22] & [24]  & [28] &[29] &[30]&[40]& $\textbf{Our work}$ \\
\hline
\hline
mmWave WSN &  &  &  &  &  & \checkmark  & \checkmark  &\checkmark & \checkmark &\checkmark\\
\hline
Vector parameter estimation & & \checkmark & \checkmark &  \checkmark  & \checkmark   & \checkmark & \checkmark &  & &\checkmark\\
\hline
Per sensor power constraint &&\checkmark & &\checkmark  & & \checkmark & \checkmark & \checkmark & \checkmark &\checkmark\\
\hline
Coherent MAC &  & \checkmark & \checkmark &  \checkmark &\checkmark  &  &\checkmark & \checkmark  & \checkmark &\checkmark\\
\hline
Hybrid transceiver design &\checkmark & & &  &   & \checkmark & \checkmark &  & \checkmark &\checkmark\\
\hline
Noiseless FC  & & &  & & &   & &  &  &\checkmark\\
\hline
Gaussian CSI uncertainty  & & &  & &  &   & &  & \checkmark &\checkmark\\
\hline
\end{tabular}
\end{table*} 
Coming next to parameter estimation in IoTNes, linear decentralized estimation (LDE), constitutes an excellent strategy \cite{xiao2008linear,behbahani2012linear} where multiple IoTNos precode their observations and transmit them to the FC using a multiple access channel (MAC) for receive combining. This in turn necessitates the design of optimal TPCs and RCs that minimize the MSE of parameter estimation at the FC. An overview of the various related works in this area is presented next. 
\subsection{Literature Review}
The most popular model of LDE has been developed in the seminal works
\cite{xiao2008linear} and \cite{behbahani2012linear}. The authors of
\cite{6949105} have proposed novel zero-forcing based TPC designs for
sub-6 GHz IoTNes for the LDE of a vector parameter. An interesting
aspect of the scheme described is that it does not require any
post-processing at the FC. The treatise \cite{9322457} developed
majorization theory-based closed-form transceiver designs considering
scenarios of both noiseless as well as noisy IoTNo observations. Zhou
\textit{et al.} \cite{7575659} described a learning-based iterative
mixed algorithm for the detection-estimation of a scalar parameter,
while also considering IoTNo defects, thus making it suitable for
practical applications. The authors of \cite{9371414} proposed an edge
computing based scheme for decentralized sensing of a spatially
correlated process. A subset of IoTNos is efficiently selected by the
algorithm, which strikes an attractive trade-off between energy
efficiency and sensing quality.

Leong \textit{et
  al.}\cite{leong2011asymptotics} conceived optimal power allocation
schemes for estimating a time varying scalar parameter. The TPCs and
RCs of \cite{leong2011asymptotics} are designed with the goal of
minimizing the resultant MSE at the FC. Furthermore, the authors of
\cite{leong2011power} analyse the MSE of estimating and tracking a
time-varying parameter. The FC of \cite{leong2011power} employs Kalman
filtering to track the parameter of interest. Akhtar \textit{et al.}
\cite{akhtar2017distributed} described TPC designs that sequentially
estimate a time-varying parameter vector. Their design is iterative in
nature and it is capable of operating in the face of time-varying
channels while obeying specific transmit power constraints. The authors
of \cite{cheng2021joint} present a framework for sequentially
estimating a parameter of interest, wherein only a subset of IoTNos
transmit information to the FC based on prior communication among the
IoTNos. A variety of schemes were proposed in
\cite{8368090,venkategowda2017data,9473656} for LDE in a wireless
powered IoTNe, where the IoTNos run on the energy harvested from the
RF signals transmitted by energy access points. A set of LDE schemes
have been proposed in~\cite{7069213,7395392,6971234} for
cutting-edge massive MIMO IoTNes, which exploit the asymptotic
beamforming properties of large antenna arrays. The authors of
\cite{liu2018secure} propose a secure diffusion-based least-mean
squares algorithm for distributed parameter estimation in the presence
of an eavesdropper. However none of the above impressive contributions
are directly applicable to mmWave MIMO systems relying on the bespoke
signal processing employed in a hybrid transceiver.  The literature
specifically related to beamforming in mmWave IoTNes is reviewed
next. \par
\begin{figure*}
\centering
\includegraphics[height=0.37\linewidth, width=0.8\linewidth]{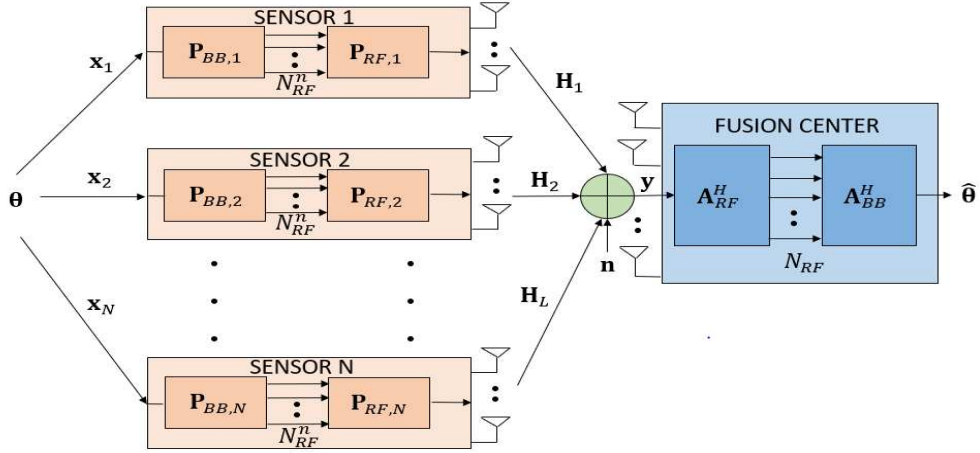}
\caption{Schematic diagram of hybrid MIMO  signal processing for vector parameter estimation in mmWave MIMO IoTNe over a coherent MAC.}
\label{fig:mmWsys_mdl}
\end{figure*}
Liu \textit{et al.} \cite{liu2020analog} developed novel hybrid transceiver designs for maximizing the spectral efficiency of an orthogonal MAC-based mmWave IoTNe. The cutting-edge scheme described initially derives the fully digital transceivers followed by decomposing them into their baseband and RF components. An iterative algorithm is proposed for obtaining the closed-form expressions of baseband TPCs. Then a pair of procedures  are proposed for the analog TPC/ RC using the popular alternating direction method of multipliers (ADMM) and the steepest descent principles, respectively. However, a drawback of the system is that it relies on an orthogonal MAC, which is not bandwidth efficient as it requires dedicated channels between the constituent IoTNos and the FC. It must be noted at this juncture that the orthogonal MAC concept considerably differs from a coherent MAC, wherein the signal received at the FC constituted by a superposition of the signals transmitted by all the IoTNos. To further develop the latter system, the authors of \cite{9314704} presented both centralized and distributed LMMSE transceiver designs for a mmWave MIMO IoTNe. A state-of-the-art hybrid transmit beamforming algorithm is proposed in \cite{8334416} for EAPs and simultaneous wireless information and power transfer (SWIPT) in the support of IoT receivers . However, a limiting aspect of the above contributions, which restricts their applicability in practice, is that they assume perfect knowledge of the channel between the FC and the constituent IoTNos. Needless to say, this is hard to achieve in practice owing to several limitations, such as the limited pilot overhead and the finite precision error-prone feedback. Papers that address this critical issue of robust TPC/ RC design, in both sub-6 GHz and mmWave systems, are discussed next.    \par 

The authors of \cite{9461735} and \cite{9464966} propose robust TPC/ RC designs for LDE in a IoTNe, for sparse and temporally correlated parameter vectors, respectively, relying on imperfect CSI. Venkategowda \textit{et al.} \cite{7807319} designed TPCs robust to imperfect CSI for estimating a scalar parameter. Their robust designs were shown to yield an improved performance in comparison to non-robust designs that are oblivious to the CSI imperfections. The treatise \cite{rostami2018precoder} proposes robust closed-form TPC matrices for parameter vector estimation. The authors of \cite{7807319}, \cite{rostami2018precoder} propose robust designs employing a zero-forcing (ZF) constraint at the FC to make the post-processing simple. However, the ZF equalizer leads to noise amplification at the receiver, which degrades the signal-to-noise ratio (SNR). Moreover, they also impose a total power constraint on the MIMO IoTNe. In the context of  mmWave MIMO systems, minimum-MSE robust transceivers are developed in \cite{9448465} for a mmWave multi-user system. Explicitly, the authors proposed a non-linear hybrid transceiver design based on Tomlinson/Harashima precoding. Jiang \textit{et al.} \cite{jiang2019mmwave} have proposed robust transceiver designs for a half-duplex mmWave relay system, where the RF TPC/ RC are designed based on the strongest eigenmode of the mmWave channel. An iterative algorithm based on mutual information maximization has also been developed in their treatise for designing the baseband TPCs. Cai  \textit{et al.} \cite{8606437} have developed a robust transceiver design based on the principle of sum-rate maximization for full-duplex (FD) MIMO relay-aided multi-user mmWave systems. Furthermore, Zhao \textit{et al.} \cite{9079551} have proposed a robust transceiver design in a full-duplex (FD) mmWave multi-cell network. In \cite{8606437}, \cite{9079551}, iterative algorithms are developed based on the classic penalty dual decomposition (PDD) technique for a scenario subject to imperfect CSI.
% The research work \cite{8606437} propounds a robust joint hybrid transceiver design for a FD MIMO relay-aided multi-user mmWave system that targets worst-case sum-rate maximization.
  The authors of \cite{9372293} characterized the performance of a robust hybrid TPC scheme in a downlink mmWave IoTNe relying on single antenna IoTNos and a multi-antenna FC. Furthermore, a noteworthy aspect of the design procedure of \cite{9372293} is that it maximizes the minimum secrecy at each IoTNo, in the presence of an eavesdropper (ED), which guarantees privacy of the transmitted information. Moreover, in their work \cite{9372293}, the CSI between the ED and the FC is assumed to be imperfectly known, whereas that between each IoTNo and the FC is assumed to be perfectly known. In the pioneering paper in \cite{RAJPUT2021101377}, the authors describe a strategy to design the hybrid TPC in a mmWave-based IoTNe with multiple-antenna IoTNos and a single antenna FC. Hybrid baseband and RF TPCs were designed that attained the minimum-MSE for LDE of the parameter. Robust hybrid TPCs were also designed accounting for the CSI uncertainty. However, the results and framework of \cite{RAJPUT2021101377} are restricted to scalar parameter estimation due to having a single antenna FC in the system. To the best of our knowledge, there is no literature on the design of the robust hybrid transceivers for the LDE of a parameter vector in a coherent MAC-based mmWave IoTNe, accounting also for realistic CSI uncertainty. We fill the knowledge-gap in the present work. Table \ref{Literature} presents a comparison of the contributions of this paper against those reviewed above.  The novel contributions of this work are boldly contrasted to the literature at a glance in Table \ref{Literature} and are also elaborated below in a point-wise fashion. \par
\begin{itemize}
\item To begin with, hybrid transceiver designs are presented for MSE minimization at the FC. The RF TPCs and RCs are determined
using the dominant array vectors of the channel. Subsequently, the BB RC is derived using the LMMSE rule, while the BB
TPC corresponding to each IoTNo is derived in 
closed-form using Cauchy's interlacing theorem \cite{xiao2008linear}.
\item For a noisy FC, the MSE minimization problem of hybrid transceiver designs is observed to be non-convex. Hence,
to render the optimization problem tractable, the non-convex constant-magnitude constraints are eliminated by initially considering the design of a fully-digital transceiver. The non-convexity of the optimization objective is handled by exploiting the  block-coordinate descent (BCD)-based iterative design procedure, which ultimately yields a closed-form expression for the optimal fully-digital transceiver in each iteration.
\item Upon convergence, the simultaneous orthogonal matching pursuit (SOMP) technique is harnessed for decomposing it into its RF and BB components.
\item To account for CSI uncertainty in real-world systems, robust transceiver designs are developed for a noisy FC scenario that mitigate the performance degradation imposed by imperfect CSI.
\item The centralised MMSE bound has been derived that benchmarks the performance of the proposed LDE schemes. Our numerical results demonstrate the effectiveness of the proposed hybrid transceiver designs, and also validate the improved performance of our robust transceiver design over its non-robust counterpart, which is oblivious of the CSI imperfection.
\end{itemize}
\subsection{Outline of the paper}
The rest of the manuscript is organized as follows. Section-\ref{Sec-II} presents our LDE system model in a mmWave MIMO IoTNe. Section-\ref{Sec-III} describes our hybrid transceiver relying on MSE minimization for a noiseless FC, while Section-\ref{Sec-IV} describes its counterpart designed for a general system relying on a noisy FC. Section-\ref{Sec-V} describes our robust hybrid transceiver designs incorporating CSI uncertainty. Finally, Section-\ref{Sec-VI} and Section-\ref{Sec-VII} present our numerical results and our conclusions, respectively.\par
\textit{Notation:} The following notation is used throughout this paper. Lowercase ($\mathbf{c}$) and uppercase ($\mathbf{C}$) letters denote vectors and matrices, respectively. The operators $(.)^T$, $(.)^H$ and $|.|$ denote the transpose, Hermitian and magnitude. $\text{Tr}(.)$ and $\mathbb{E}[.]$ represent the trace and statistical expectation operators, respectively. For a matrix $\mathbf{C}$, its $(i,j)$th element is denoted as $\mathbf{C}(i,j)$, while its column and row spaces are denoted by $\mathcal{C}(\mathbf{C})$ and $\mathcal{R}(\mathbf{C})$, respectively. The Frobenius and $l_0$ norms are represented as $||.||_F$ and $||.||_0$. Finally, $\mathbf{c}=\text{diag}(\mathbf{C})$ denotes a vector comprising of the principal diagonal elements of the matrix $\mathbf{C}$.
 
\section{mmWave WSN System and Channel Model }\label{Sec-II}
A mmWave-based MIMO IoTNe supporting $N$ IoTNos and a FC is considered as shown in Fig. \ref{fig:mmWsys_mdl}. Each IoTNo employs $N_t$ transmit antennas (TAs) and $N_{\text{RF}}^n$ RF chains, while the FC has $N_r$ receive antennas (RAs) and $N_{\text{RF}}$ RF chains, where $N_{\text{RF}}^n, N_{\text{RF}} << \text{min}(N_t,N_r)$. The IoTNos are sensing/ monitoring a common parameter vector $\boldsymbol{{\theta}}\in \mathbb{C}^{q \times 1}$ comprised of $q$ different physical quantities that have to be estimated.
%For example, in a smart agriculture application, the different quantities of interest may be temperature, soil moisture, pressure etc..
The observation vector of IoTNo $n$, represented by ${\mathbf{x}}_{n} \in \mathbb{C}^{l \times 1}$ can be modeled as
\begin{align} \label{eq:BB}
{\mathbf{x}}_{n}={\mathbf{C}}_{n}\boldsymbol{\theta}+{\mathbf{v}}_{n},
\end{align}
where $\mathbf{C}_n \in \mathbb{C}^{l \times q}$ and $\mathbf{v}_n \sim \mathcal{CN}(\mathbf{0},\mathbf{R}_n)\in \mathbb{C}^{l \times 1}$ represent the observation matrix  and the observation noise corresponding to the $n$th IoTNo, respectively. Each node employs the hybrid TPC comprised of the cascaded BB and RF TPC matrices $\mathbf{P}_{\text{BB},n} \in { {\mathbb {C}}}^{ N_{\text{RF}}^n \times l}$ and  $\mathbf{P}_{\text{RF},n} \in \mathbb{C}^{N_t \times N_{\text{RF}}^n}$, respectively. Subsequently, each IoTNo transmits its precoded observations to the FC over a coherent multiple access channel (MAC). Therefore, the signal $\mathbf{y}  \in { {\mathbb {C}}}^{N_r \times 1}$ received at the FC can be modeled as
\begin{align} \label{eq6}
{ \mathbf{y} } & =\sum_{n=1}^{N} { \mathbf{H} }_{n }{ \mathbf{P} }_{ \text{RF},n }{\mathbf{P}}_{ \text{BB},n }{ \mathbf{x} }_{ n } + \mathbf{w} \nonumber \\
& =  \sum_{n=1}^{N} { \mathbf{H} }_{ n }{ \mathbf{P} }_{ \text{RF},n }{\mathbf{P}}_{ \text{BB},n }{ \mathbf{C} }_{ n }\boldsymbol{\theta} +\sum_{n=1}^{N}{ \mathbf{H} }_{ n }{ \mathbf{P} }_{ \text{RF},n }{\mathbf{P}}_{ \text{BB},n }{ \mathbf{v} }_{ n }+{\mathbf{w}} \nonumber \\
& = { \mathbf{H} }\mathbf{P}_{\text{RF}}\mathbf{P}_{\text{BB}}{ \mathbf{C} } \boldsymbol{\theta} + { \mathbf{H} }\mathbf{P}_{\text{RF}}\mathbf{P}_{\text{BB}} \mathbf{v}+{\mathbf{w}}, 
\end{align}
where the matrix $\mathbf{H}_n \in {\mathbb{C}}^{N_r \times N_t}$ represents the MIMO channel matrix between each IoTNo $n$ and the FC, whereas the vector $\mathbf{w}\in {\mathbb{C}}^{N_r \times 1}$ represents the FC noise and follows the distribution $\mathcal{CN}(\mathbf{0},\mathbf{R}_w)$. The stacked observation matrix $\mathbf{C} \in {\mathbb{C}}^{Nl \times q}$, concatenated channel matrix $\mathbf{H} \in {\mathbb{C}}^{N_r\times NN_t}$, the overall block-diagonal BB TPC $\mathbf{P}_{\text{BB}} \in { {\mathbb {C}}}^{ NN^n_{\text{RF}} \times Nl}$ and RF TPC $\mathbf{P}_{\text{RF}} \in { {\mathbb {C}}}^{ NN_{t} \times NN^n_{\text{RF}}}$ are defined as
\begin{align}
\mathbf{C} &= [\mathbf{C}_1^T,\mathbf{C}_2^T,\hdots,\mathbf{C}_N^T]^T,\\ 
\mathbf{H} &= [\mathbf{H}_1,\mathbf{H}_2,\hdots,\mathbf{H}_N],\\
\mathbf{P}_{\text{BB}}&= \mathrm{diag} (\mathbf{P}_{\text{BB},1},\hdots,\mathbf{P}_{\text{BB},N}), \\
\mathbf{P}_{\text{RF}}&= \mathrm{diag} (\mathbf{P}_{\text{RF},1},\hdots,\mathbf{P}_{\text{RF},N}).
\end{align}
At the FC, the hybrid RC operation is performed using the cascaded RF and BB combining matrices $\mathbf{A}_{\text{RF}} \in \mathbb{C}^{N_r \times N_{\text{RF}}}$ and $\mathbf{A}_{\text{BB}} \in \mathbb{C}^{N_{\text{RF}}\times q}$, respectively, to yield the estimate $\widehat{\boldsymbol{\theta}}$ as
\begin{align}
\widehat{\boldsymbol{\theta}}=\mathbf{A}^H_{\text{RF}}\mathbf{A}^H_{\text{BB}}\mathbf{y}.
\end{align}
Since, the hybrid MIMO architecture employs the constant magnitude RF TPC and RC matrices, their elements can be constrained as follows  
\begin{align}
&\left \vert \mathbf{P}_{\text{RF},n}(i,j)\right\vert = \frac{1}{\sqrt{N_t}}, \; 1 \leq n \leq N,\; \forall\; i,j,\nonumber \\
&\left \vert \mathbf{A}_{\text{RF}}(i,j)\right\vert = \frac{1}{\sqrt{N_r}},\; \forall\; i,j.
\end{align}
%\subsection{mmWave MIMO Channel Model}
Furthermore, the MIMO channel matrix $\mathbf {H}_n$ is defined as \cite{tse2005fundamentals,
heath2016overview} 
\begin{align}
\mathbf{H}_n=\sqrt{\displaystyle\frac{N_rN_t}{K}}\sum\limits_{k=1}^{K}\alpha_{k,n} \mathbf{a}_{\text{FC}}(\phi_k) \mathbf{a}^H_{\text{s}}(\theta_{k,n}),
\end{align}
where the 3-tuple $(\alpha_{k,n}, \phi_k, \theta_{k,n})$ represents the complex gain $\alpha_{k,n}$, angle of arrival (AoA) $\phi_k$ at the FC, and angle of departure (AoD) $\theta_{k,n}$ at the $n$th IoTNe associated with the $k$th cluster and $K$ denotes the total number of clusters. The vectors $\mathbf{a}_{\text{FC}}(\phi_k)\in \mathbb{C}^{N_r \times 1}$ and $\mathbf{a}_{\text{s}}(\theta_{k,n})\in \mathbb{C}^{N_t \times 1}$ represent the array response vectors at the FC and the $n$th IoTNo, respectively, corresponding to the $k$th cluster, which are defined as
\begin{equation}
\mathbf{a}_{\text{FC}}(\phi_k)=\hspace{-2pt}\displaystyle\frac{1}{\sqrt{N_R}} \Big[ 1, e^{-j\tilde{\phi}_k}, \cdots , e^{-j(r-1)\tilde{\phi}_k}\Big]^T\hspace{-5pt}, \label{eq: rxarrresvec}
\end{equation}
\begin{equation}
\mathbf{a}_{\text{s}}(\theta_{k,i})=\hspace{-2pt}\displaystyle\frac{1}{\sqrt{N_T}} \Big[ 1, e^{-j\tilde{\theta}_{k,n}}, \cdots , e^{-j(t-1)\tilde{\theta}_{k,n}}\Big]^T\hspace{-4pt}, \label{eq: txarrresvec}
\end{equation}
where $\tilde{\phi}_k = \frac{2 \pi}{\lambda}d_R \cos \phi_k$ and $\tilde{\theta}_{k,n}= \frac{2 \pi}{\lambda}d_T \cos \theta_{k,n}$. The quantities $\lambda, d_R,$ and $d_T$ denote the carrier's wavelength and inter-antenna spacings at the FC and each IoTNo, respectively. The mmWave MIMO channel $\mathbf{H}_n$ can be equivalently expressed in the compact form of
\begin{align}\label{Channel:EQ}
\mathbf{H}_n = \mathbf{A}_{\text{FC}} \mathbf{D}_n \mathbf{A}_{\text{s},n}^{H},
\end{align}
where $\mathbf{A}_{\text{FC}} = [\mathbf a_{\text{FC}}(\phi_1), \cdots  , \mathbf a_{\text{FC}}(\phi_{K})] \in \mathbb{C}^{N_r \times K} , \ \mathbf{A}_{\text{s},n} = [\mathbf a_{\text{s}}(\theta_{1,n}), \cdots  , \mathbf a_{\text{s}}(\theta_{K,n})] \in \mathbb{C}^{N_t \times K}$ are termed the array response matrices corresponding to the FC and the $n$th IoTNo, respectively, and $\mathbf{D}_n =\sqrt{\frac{N_rN_t}{K}} \mathrm{diag}(\alpha_{1,n}, \cdots, \alpha_{K,n})$ comprises of the path-gains. We assume that the FC has perfect knowledge of the array response matrices $\mathbf{A}_{\text{FC}}$, $\mathbf{A}_{\text{s},n}$ and the gain matrix $\mathbf{D}_n$. The next section describes our novel hybrid transceiver designed for the MSE minimization of a noiseless FC. 
\begin{figure} 
\begin{center}
\includegraphics[width=\linewidth]{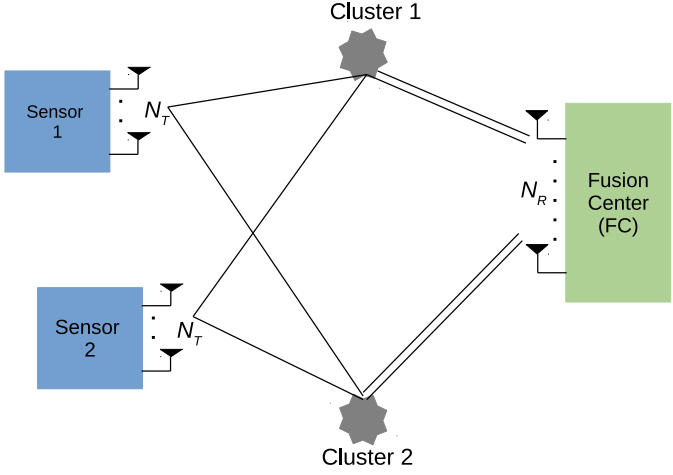}
\end{center} 
\caption{Spatial mmWave MIMO channel model}
\label{fig:mmWsys_mdl1}
\end{figure}
\section{Hybrid Transceiver Design for a Noiseless FC}\label{Sec-III}
The purpose of this study is to determine the minimum number of RF chains required at each sensor to achieve the minimum MSE. We would like to emphasize that, in this scenario, the observation noise dominates the thermal noise at the receiver, which renders the latter negligible. Thus, the measurements of each sensor are corrupted by observation noise, which necessitates the design of the  hybrid TPCs. On account of a noiseless FC, the received vector $\mathbf{y}$ can be obtained by substituting $\mathbf{w}=\mathbf{0}$ in \eqref{eq6}. The estimate $\widehat{\boldsymbol\theta} \in {\mathbb{C}}^{m \times 1}$ at the output of the LMMSE RC for a noiseless coherent MAC is given by
\begin{align}
 \widehat{\boldsymbol\theta} &= \mathbf{A}_{\text{BB}}^H \mathbf{A}_{\text{RF}}^H \mathbf{y}\nonumber\\
 &=\mathbf{A}_{\text{BB}}^H \underbrace{\mathbf{A}_{\text{RF}}^H { \mathbf{H} }\mathbf{P}_{\text{RF}}}_{\bar{\mathbf{H}}}\mathbf{P}_{\text{BB}}{ \mathbf{C} } \boldsymbol{\theta} + \mathbf{A}_{\text{BB}}^H \mathbf{A}_{\text{RF}}^H { \mathbf{H} }\mathbf{P}_{\text{RF}}\mathbf{P}_{\text{BB}} \mathbf{v}. \label{eq: compact_sys_mdl}
\end{align}
Furthermore, the resultant MSE at the FC, which is defined as $\eta\left(\mathbf{P}_{\text{RF}},\mathbf{P}_{\text{BB}}, \mathbf{A}_{\text{RF}}\right)= \mathbb{E}\left[ \left\Vert \widehat{\boldsymbol\theta} - \boldsymbol\theta \right\Vert^2\right]$, can be expressed as
\begin{align}
&\eta\left(\mathbf{P}_{\text{RF}},\mathbf{P}_{\text{BB}}, \mathbf{A}_{\text{RF}}\right) = \mathrm{Tr}\Big[\sigma_{\theta}^2 \mathbf{I}_{m} + \mathbf{C}^H \mathbf{P}_{\text{BB}}^H \mathbf{P}_{\text{RF}}^H{\bar{\mathbf{H}}}^H \bigg(\mathbf{A}_{\text{RF}}{\bar{\mathbf{H}}} \nonumber \\
&\mathbf{P}_{\text{RF}}\mathbf{P}_{\text{BB}} \mathbf{R}_{v}\mathbf{P}_{\text{BB}}^H\mathbf{P}^H_{\text{RF}}{\bar{\mathbf{H}}}^H\mathbf{A}_{\text{RF}} \bigg)^{-1}\mathbf{A}^H_{\text{RF}}{\bar{\mathbf{H}}}\mathbf{P}_{\text{RF}} \mathbf{P}_{\text{BB}}  \mathbf{C} \Big]^{-1}.\label{eq:mse_theta}
\end{align} 
Hence, we can formulate the optimization problem of our hybrid transceiver with an objective to minimize the MSE in \eqref{eq:mse_theta} as follows
\begin{align}
&\underset{\mathbf{P}_{\text{RF}},\mathbf{P}_{\text{BB}}, \mathbf{A}_{\text{RF}}}{\text{minimize}}\quad \eta\left(\mathbf{P}_{\text{RF}},\mathbf{P}_{\text{BB}}, \mathbf{A}_{\text{RF}}\right)
\nonumber\\
& \text{subject to } 
\left \vert \mathbf{P}_{\text{RF},n}(i,j)\right\vert = \frac{1}{\sqrt{N_t}},1 \leq n \leq N,\forall\; i,j,\nonumber \\
& \;\; \;\; \;\; \;\; \;\; \;\; \;\; \;\ \left \vert \mathbf{A}_{\text{RF}}(i,j)\right\vert = \frac{1}{\sqrt{N_r}},\; \forall\; i,j        . \label{eq:gen_opt_prob}
\end{align}
The constant magnitude of the elements in the RF TPC $\mathbf{P}_{\text{RF},n}$ and RC $\mathbf{A}_{\text{RF}}$ results in the non-convexity of the above optimization problem, and makes it intractable.
% However, as shown in the following sequel, for the given RF precoders and combiner, the resultant optimization problem has a closed form solution.
In this context, suitable RF TPCs $\mathbf{P}_{\text{RF},n}$ and RC $\mathbf{A}_{\text{RF}}$ can be obtained by using the procedure outlined in the next subsection, which is followed by the design of the optimal BB TPC $\mathbf{P}_{\text{BB}}$ for each IoTNo $n$. This section considers the variance of each parameter to be $\sigma_{\theta}^2 = 1$ and the observation noise covariance to be $\mathbf{R}_{n}=\mathbf{I}_{l}$. 
\subsection{Design of RF TPCs $\mathbf{P}_{\text{RF},n}$ and RC $\mathbf{A}_{\text{RF}}$}
Since the columns of the matrices $\mathbf{A}_{\text{FC}}$ and $\mathbf{A}_{\text{s},n}$ are the array response vectors corresponding to the AoAs at the FC and AoDs at the $n$th IoTNo, the $N_{\text{RF}}$ columns of the RF RC $\mathbf{A}_{\text{RF}}$ and TPC $\mathbf{P}_{\text{RF},n}$ can be set using the $N_{\text{RF}}$ dominant array response vectors, respectively. In order to design the RF TPC for the $n$th IoTNo, we organise the complex gains  $\alpha_{k,n}$ in decreasing order of magnitude so that $\vert\alpha_{k_1,n}\vert \geq \vert\alpha_{k_2,n}\vert \geq  \cdots \geq \vert\alpha_{k_{K},n}\vert$. The RF TPC $\mathbf{P}_{\text{RF},n}$ for the $n$th IoTNo can be designed by selecting the $N_{\text{RF}}$ columns denoted by the set $\mathcal{K}_{n} = \left\{ k_{1}, k_{2}, \hdots, k_{N_{\text{RF}}^n} \right\}$, which contains the indices of the $N_{\text{RF}}$ dominant transmit array response vectors. This is mathematically represented as
\begin{align}
\mathbf{P}_{\text{RF},n} = \mathbf{A}_{\text{s},n}\left( \ :\ ,\  \mathcal{K}_n\ \right).
\end{align} 
The design of the RF RC $\mathbf{A}_{\text{RF}}$ is described next. Let the quantity $\alpha_k = \sum_{n=1}^{N} \vert \alpha_{k,n} \vert$ denote the sum of the magnitudes of the complex path-gains of each IoTNo corresponding to the $k$th cluster. One can now arrange the quantities $\alpha_{l}$ in decreasing order of their magnitude, so that $\vert\alpha_{k_1}\vert \geq \vert\alpha_{k_2}\vert \geq  \cdots \geq \vert\alpha_{k_N}\vert$. The RF RC $\mathbf{A}_{\text{RF}}$ is obtained as
\begin{align}
\mathbf{A}_{\text{RF}} = \mathbf{A}_{\text{FC}}\left( \ :\ ,\  \mathcal{K}\ \right),
\end{align}
where we have $\mathcal{K} = \left\{ k_{1}, k_{2}, \hdots, k_{N_{\text{RF}}} \right\}$. Employing these settings of the RF TPC $\mathbf{P}_{\text{RF},n}$ and RC $\mathbf{A}_{\text{RF}}$, the design of the optimal BB TPC $\mathbf{P}_{\text{BB},n}$ for each IoTNo $n$ is described next. 
\subsection{Baseband TPC $\mathbf{P}_{\text{BB},n}$ design}
Upon employing the LMMSE BB RC $\mathbf{A}_{\text{BB}}$ at the FC as shown in \eqref{eq: compact_sys_mdl}, the resultant error covariance matrix $\mathbf{E}$ is given by 
\begin{align}
\mathbf{E}=\left[\mathbf{I}_q+\mathbf{C}^H\widetilde{\mathbf{H}}^H\left(\widetilde{\mathbf{H}}\widetilde{\mathbf{H}}^H\right)^{-1}\widetilde{\mathbf{H}}\mathbf{C}\right]^{-1},
\end{align}
where we have $\widetilde{\mathbf{H}}=\bar{\mathbf{H}}\mathbf{P}_{\text{BB}} \in \mathbb{C}^{N_{\text{RF}}^n \times Nl}$. Furthermore, assuming that the effective channel $\bar{\mathbf{H}}_n = \mathbf{A}_{\text{RF}}^H { \mathbf{H}_n }\mathbf{P}_{\text{RF},n} \in \mathbb{C}^{N_{\text{RF}}^n \times l}$ between each IoTNo $n$ and the FC is invertible, solving the optimization problem for $\mathbf{P}_{\text{BB},n}$ is equivalent to solving it for $\widetilde{\mathbf{H}}_n = \bar{\mathbf{H}}_n \mathbf{P}_{\text{BB},n}$. As a result, the optimization problem in \eqref{eq:gen_opt_prob} can be restated as
 \begin{align}
\underset{\left\{\tilde{\mathbf{H}}_{n}\right\}_{n=1}^N}{\text{minimize }}  &\mathrm{Tr}\left(\mathbf{E}\right) \nonumber \\
\text{subject to } &\mathbf{E}^{-1}=\left[ \mathbf{I}_{q} + \mathbf{C}^H \widetilde{\mathbf{H}}^H \left(\widetilde{\mathbf{H}}\widetilde{\mathbf{H}}^H \right)^{-1}\widetilde{\mathbf{H}} \mathbf{C} \right]. \label{eq:FB_opt}
\end{align}
The above problem can be solved using Cauchy's interlacing theorem which is described next in detail. Let us assume that the singular value decomposition (SVD) of $\widetilde{\mathbf{H}}$ is given by 
\begin{align}
\widetilde{\mathbf{H}}=\mathbf{U}_{\tilde{H}} \boldsymbol{\Sigma}_{\tilde{H}}\mathbf{V}^H_{\tilde{H}} ,
\end{align}
where $\mathbf{U}_{\tilde{H}}  \in \mathbb{C}^{N_{\text{RF}}^n \times N_{\text{RF}}^n}$, $\mathbf{V}_{\tilde{H}}  \in \mathbb{C}^{Nl \times Nl}$ are unitary matrices and $\boldsymbol{\Sigma}_{\tilde{H}}  \in \mathbb{C}^{N_{\text{RF}}^n \times Nl}$ is the matrix containing singular values on its principal diagonal. Now, substituting the SVD of  $\widetilde{\mathbf{H}}$ into $\widetilde{\mathbf{H}}^H \left(\widetilde{\mathbf{H}}\widetilde{\mathbf{H}}^H \right)^{-1}\widetilde{\mathbf{H}}$, one obtains
\begin{align}
&\widetilde{\mathbf{H}}^H \left(\widetilde{\mathbf{H}}\widetilde{\mathbf{H}}^H \right)^{-1}\widetilde{\mathbf{H}}
%&=\mathbf{V}_{\tilde{H}}\boldsymbol{\Sigma}^H_{\tilde{H}}\mathbf{U}^H_{\tilde{H}}\left( \mathbf{U}_{\tilde{H}} \boldsymbol{\Sigma}_{\tilde{H}}\boldsymbol{\Sigma}^H_{\tilde{H}}\mathbf{U}^H_{\tilde{H}}\right)^{-1}\mathbf{U}_{\tilde{H}} \boldsymbol{\Sigma}_{\tilde{H}}\mathbf{V}^H_{\tilde{H}} \nonumber \\
%&= \mathbf{V}_{\tilde{H}}\boldsymbol{\Sigma}^H_{\tilde{H}}\left(\boldsymbol{\Sigma}_{\tilde{H}}\boldsymbol{\Sigma}^H_{\tilde{H}} \right)\boldsymbol{\Sigma}_{\tilde{H}}\mathbf{V}^H_{\tilde{H}} \nonumber \\
=\mathbf{V}_{\tilde{H}} 
\begin{bmatrix}
\mathbf{I}_r & \mathbf{0}  \\
\mathbf{0} & \mathbf{0}  
\end{bmatrix}
\mathbf{V}^H_{\tilde{H}}, \label{eq:1}
\end{align}
where we have $r = \mathrm{rank}\left( \widetilde{\mathbf{H}} \right)$.
%then the term $\mathbf{G}^H\widetilde{\mathbf{H}}^H \left(\widetilde{\mathbf{H}}\widetilde{\mathbf{H}}^H \right)^{-1}\widetilde{\mathbf{H}}\mathbf{G}$, can be further simplified as shown next
%\begin{align}
%&\mathbf{G}^H\widetilde{\mathbf{H}}^H \left(\widetilde{\mathbf{H}}\widetilde{\mathbf{H}}^H \right)^{-1}\widetilde{\mathbf{H}}\mathbf{G} \nonumber \\
%&=\mathbf{V}_{G} \boldsymbol{\Sigma}^H_{G} \mathbf{U}^H_{G} \mathbf{V}_{\tilde{H}} 
%\begin{bmatrix}
%\mathbf{I}_r & \mathbf{0}  \\
%\mathbf{0} & \mathbf{0}  
%\end{bmatrix}
%\mathbf{V}^H_{\tilde{H}}\mathbf{U}_{G} \boldsymbol{\Sigma}_{G}\mathbf{V}^H_{G}.\label{eq:2}
%\end{align} 
Upon substituting the SVD of $\mathbf{C}$ as $\mathbf{C}=\mathbf{U}_{C} \boldsymbol{\Sigma}_{C}\mathbf{V}^H_{C}$ into $\mathbf{V}^H_{C}\mathbf{E}^{-1}\mathbf{V}_{C}$, we arrive at:
\begin{align}
\mathbf{V}^H_{C}\mathbf{E}^{-1}\mathbf{V}_{C}&=\mathbf{V}^H_{C}\left( \mathbf{I}_{q} + \mathbf{C}^H \widetilde{\mathbf{H}}^H \left(\widetilde{\mathbf{H}}\widetilde{\mathbf{H}}^H \right)^{-1}\widetilde{\mathbf{H}} \mathbf{C} \right)\mathbf{V}_{C} \nonumber \\
& =\mathbf{I}_{q}+\boldsymbol{\Sigma}^H_{C} \mathbf{U}^H_{C} \mathbf{V}_{\tilde{H}} 
\begin{bmatrix}
\mathbf{I}_r & \mathbf{0}  \\
\mathbf{0} & \mathbf{0}  
\end{bmatrix}
\mathbf{V}^H_{\tilde{H}}\mathbf{U}_{C} \boldsymbol{\Sigma}_{C} \nonumber \\
& =\mathbf{I}_{q}+\boldsymbol{\Sigma}^H_{C} \mathbf{Q}^H 
\begin{bmatrix}
\mathbf{I}_r & \mathbf{0}  \\
\mathbf{0} & \mathbf{0}  
\end{bmatrix}
\mathbf{Q} \boldsymbol{\Sigma}_{C}, \label{eq:3}
\end{align}
where the unitary matrix obeys $\mathbf{Q}=\mathbf{V}^H_{\tilde{H}}\mathbf{U}_{C} \in \mathbb{C}^{Nl \times Nl}$. Let $\lambda_k\left(\mathbf{X}\right)$ denote the $k$th smallest eigenvalue of any symmetric matrix $\mathbf{X}$. Then it follows from \eqref{eq:3} that
\begin{align}
\lambda_k\left(\mathbf{E}^{-1}\right)&=\lambda_k\left(\mathbf{V}^H_{C}\mathbf{E}^{-1}\mathbf{V}_{C}\right)\nonumber \\
&=\lambda_k\left(\mathbf{I}_{q}+\boldsymbol{\Sigma}^H_{C} \mathbf{Q}^H 
\begin{bmatrix}
\mathbf{I}_r & \mathbf{0}  \\
\mathbf{0} & \mathbf{0}  
\end{bmatrix}
\mathbf{Q} \boldsymbol{\Sigma}_{C}\right)\nonumber \\
&=\lambda_k\left(\mathbf{I}_{Nl}+\begin{bmatrix}
\mathbf{I}_r & \mathbf{0}  \\
\mathbf{0} & \mathbf{0}  
\end{bmatrix}\mathbf{Q} \boldsymbol{\Sigma}_{C}\boldsymbol{\Sigma}^H_{C} \mathbf{Q}^H 
\begin{bmatrix}
\mathbf{I}_r & \mathbf{0}  \\
\mathbf{0} & \mathbf{0}  
\end{bmatrix}
\right)\nonumber \\
&=\lambda_k\left(\mathbf{I}_{r}+\left(\mathbf{Q} \boldsymbol{\Sigma}_{C}\boldsymbol{\Sigma}^H_{C} \mathbf{Q}^H \right)_{(1:r,1:r)}\right), \label{eq:4}
\end{align}
where $\left(\mathbf{Q} \boldsymbol{\Sigma}_{C}\boldsymbol{\Sigma}^H_{C} \mathbf{Q}^H \right)_{(1:r,1:r)}$ represents the upper $r \times r$ block of the matrix $\mathbf{Q} \boldsymbol{\Sigma}_{C}\boldsymbol{\Sigma}^H_{C} \mathbf{Q}^H$. By exploiting Cauchy's interlacing theorem for the eigenvalues of a subblock matrix, we have
\begin{align}
\lambda_k\left(\left(\mathbf{Q} \boldsymbol{\Sigma}_{C}\boldsymbol{\Sigma}^H_{C} \mathbf{Q}^H \right)_{(1:r,1:r)}\right) \leq \lambda_k \left(\boldsymbol{\Sigma}_{C}\boldsymbol{\Sigma}^H_{C}\right)=\lambda_k\left(\mathbf{C}\mathbf{C}^H \right), \label{eq:5}
\end{align} 
for $1 \leq k \leq r$. Therefore, upon using \eqref{eq:4} and \eqref{eq:5}, we may conclude that 
\begin{align}
\lambda_k\left(\mathbf{E}^{-1}\right)&=\lambda_k\left(\mathbf{I}_{r}+\left(\mathbf{Q} \boldsymbol{\Sigma}_{C}\boldsymbol{\Sigma}^H_{C} \mathbf{Q}^H \right)_{(1:r,1:r)}\right) \nonumber \\
&\leq 1+\lambda_k\left(\mathbf{C}\mathbf{C}^H\right), \text{for}\; 1 \leq k \leq r
\end{align}
and $\lambda_k\left(\mathbf{E}^{-1}\right)=1$, when $r < k \leq m$. Hence, the MSE can be lower-bounded as
\begin{align}
\mathrm{MSE}=\mathrm{Tr}\left(\mathbf{E}\right)&=\sum_{k=1}^{q}\lambda_k\left(\mathbf{E}\right) \nonumber \\
&\geq \sum_{k=1}^{r}\frac{1}{1+\lambda_k\left(\mathbf{C}\mathbf{C}^H\right)}+\left(q-r\right). \label{eq:6}
\end{align}
It follows from the above relationship that the minimum MSE is obtained when $q=r$. Furthermore, observe that $r = \mathrm{rank}\left( \widetilde{\mathbf{H}} \right) \leq N_{\text{RF}}^n$, hence the lower bound is achieved when $N_{\text{RF}}^n \geq q$, i.e., the minimum number of RF chains required is equal to the parameter dimension. This implies that the singular matrix $\boldsymbol{\Sigma}_{\tilde{H}}$ has the following structure
\begin{align}
\boldsymbol{\Sigma}_{\tilde{H}}=\begin{bmatrix}
\boldsymbol{\Sigma}_{N_{\text{RF}}^n \times N_{\text{RF}}^n} & \mathbf{0}_{N_{\text{RF}}^n \times \left(Nl-N_{\text{RF}}^n\right)}   
\end{bmatrix}. \label{eq:7}
\end{align}
Furthermore, in order to achieve the lower bound in \eqref{eq:6}, the inequality in \eqref{eq:5} should be satisfied with equality. This implies that the matrix $\mathbf{Q}$ should be set as an identity matrix of dimension $N_{\text{RF}}^n \times N_{\text{RF}}^n$, which results in the relationship $\mathbf{V}_{\tilde{H}}=\mathbf{U}_C$. Hence, the optimal matrix $\widetilde{\mathbf{H}}$, denoted by $\widetilde{\mathbf{H}}_{\text{opt}}$, can be set as
\begin{align}
\widetilde{\mathbf{H}}_{\text{opt}}=\mathbf{U}_{\tilde{H}}\boldsymbol{\Sigma}_{\tilde{H}}\mathbf{U}^H_C,
\end{align} 
where $\mathbf{U}_{\tilde{H}}$ is any unitary matrix, since it does affect the MSE. Subsequently, the baseband precoder matrix for the individual sensors can be extracted using
\begin{align}
\mathbf{P}_{\text{BB},n}=\bar{\mathbf{H}}^{-1}_n \widetilde{\mathbf{H}}_{n,\text{opt}},
\end{align}
where $\widetilde{\mathbf{H}}_{n,\text{opt}} = \widetilde{\mathbf{H}}_{\text{opt}} (\ :\ ,(n-1)q+1:nq)$.
Finally, the relationship between the MSE performance achieved and the number of RF chains used can be summarized as follows:
\[
    \mathrm{MSE}= 
\begin{cases}
    \left(q-N_{\text{RF}}^n\right)+\sum_{k=1}^{N_{\text{RF}}^n}\frac{1}{1+\lambda_k\left(\mathbf{C}\mathbf{C}^H\right)},& N_{\text{RF}}^n < q,\\
    \sum_{k=1}^{q}\frac{1}{1+\lambda_k\left(\mathbf{C}\mathbf{C}^H\right)},              & N_{\text{RF}}^n \geq q.
\end{cases}
\]
The next section considers the most general scenario, where both realistic observation and FC noise are present and develops the hybrid minimum-MSE transceiver design, while also taking the available transmit power at each IoTNo into account. 
\begin{figure*}
\begin{align}
\zeta\left(\mathbf{A}_{\text{RF}}, \mathbf{A}_{\text{BB}}, \mathbf{P}_{\text{RF},n}, \mathbf{P}_{\text{BB},n}\right)=&\text{Tr}\Bigg[\mathbf{A}\left(\sum_{n=1}^{N}\mathbf{H}_n\mathbf{P}_n\left(\mathbf{C}_n\mathbf{R}_{\theta}\mathbf{C}_n^H+\mathbf{R}_n\right)\mathbf{P}^H_n\mathbf{H}^H_n\right)\mathbf{A}^H+\mathbf{A}\left(\sum_{n=1}^{N}\sum_{j=1,j\neq n}^{N}\mathbf{H}_n\mathbf{P}_n\mathbf{C}_n\mathbf{R}_{\theta}\mathbf{C}^H_j\mathbf{P}^H_j\mathbf{H}^H_j\right) \nonumber \\
&\mathbf{A}^H-\mathbf{A}\left(\sum_{n=1}^N \mathbf{H}_n\mathbf{P}_n \mathbf{C}_n\right)\mathbf{R}_{\theta}-\mathbf{R}_{\theta}\left(\sum_{n=1}^N\mathbf{C}^H_n\mathbf{P}^H_n\mathbf{H}^H_n\right)\mathbf{A}^H+\mathbf{A}\mathbf{R}_w\mathbf{A}^H+\mathbf{R}_{\theta}\Bigg].\label{MSE_2}
\end{align}
\hrulefill
\end{figure*}
\section{Hybrid Transceiver Design for a Noisy FC}\label{Sec-IV}
Again, this section develops a novel hybrid transceiver design for the more general scenario, where realistic channel-noise is also present. Additionally, another challenging aspect is also considered, where one does not have perfect knowledge of the dominant array response matrices. Hence, in this scenario, initially an iterative algorithm is developed for designing the minimum-MSE optimal fully-digital TPCs and RC matrices. Once convergence is achieved, the fully-digital matrices are decomposed into their RF and baseband counterparts using the popular SOMP algorithm. The detailed procedure is described next. The estimate $\widehat{\boldsymbol{\theta}}$ for the received vector $\mathbf{y}$ in \eqref{eq6} can be obtained as
\begin{align}\label{theta_est}
\widehat{\boldsymbol{\theta}}=& \mathbf{A}_{\text{BB}}^H \mathbf{A}_{\text{RF}}^H \mathbf{y}\nonumber\\
 =&\mathbf{A}_{\text{BB}}^H \mathbf{A}_{\text{RF}}^H \sum_{n=1}^{N} { \mathbf{H} }_{ n }{ \mathbf{P} }_{ \text{RF},n }{\mathbf{P}}_{ \text{BB},n }{ \mathbf{C} }_{ n }\boldsymbol{\theta} +\mathbf{A}_{\text{BB}}^H \mathbf{A}_{\text{RF}}^H \times \nonumber \\
 &\sum_{n=1}^{N}{ \mathbf{H} }_{ n }{ \mathbf{P} }_{ \text{RF},n }{\mathbf{P}}_{ \text{BB},n }{ \mathbf{v} }_{ n }+\mathbf{A}_{\text{BB}}^H \mathbf{A}_{\text{RF}}^H {\mathbf{w}}.
\end{align}
The expression of the resultant MSE, which denoted by $\zeta\left(\mathbf{A}_{\text{RF}}, \mathbf{A}_{\text{BB}}, \mathbf{P}_{\text{RF},n}, \mathbf{P}_{\text{BB},n}\right)$, is given in \eqref{MSE_2}. The average transmit power corresponding to the $n$th IoTNo can be evaluated as
\begin{align}\label{Pow_Cons}
&\mathbb{E}\left[||\mathbf{P}_{\text{RF},n}\mathbf{P}_{\text{BB},n}\mathbf{x}_n||^2\right]\nonumber \\
&=\text{Tr}\left[\mathbf{P}_{\text{RF},n}\mathbf{P}_{\text{BB},n}\left(\mathbf{C}_n\mathbf{R}_{\boldsymbol{\theta}}\mathbf{C}^H_n+\mathbf{R}_n\right)\mathbf{P}^H_{\text{BB},n}\mathbf{P}^H_{\text{RF},n}\right]\leq \rho_n,
\end{align}
where $\rho_n$ represents the maximum available transmit power at the $n$th IoTNo. Hence, the MSE-minimization optimization problem for designing the hybrid transceiver can be formulated as
\begin{align}\label{EQ:opt_NFC}
&{\text{minimize}} \quad \zeta\left(\mathbf{A}_{\text{RF}}, \mathbf{A}_{\text{BB}}, \mathbf{P}_{\text{RF},n}, \mathbf{P}_{\text{BB},n}\right) \nonumber\\
&\text{subject to } 
\left \vert \mathbf{P}_{\text{RF}}(i,j)\right\vert = \frac{1}{\sqrt{N_t}}, \left \vert \mathbf{A}_{\text{RF}}(i,j)\right\vert = \frac{1}{\sqrt{N_r}},\; \forall\; i,j\nonumber \\
&  \text{Tr}\left[\mathbf{P}_{\text{RF},n}\mathbf{P}_{\text{BB},n}\left(\mathbf{C}_n\mathbf{R}_{\boldsymbol{\theta}}\mathbf{C}^H_n+\mathbf{R}_n\right)\mathbf{P}^H_{\text{BB},n}\mathbf{P}^H_{\text{RF},n}\right]\leq \rho_n.
\end{align}
The above optimization problem is once again non-convex due to the constant-magnitude constraint as well as the non-convex objective function. However, for designing the fully-digital transceiver, one can ignore the non-convex constant magnitude constraint. Furthermore, the block-convex nature of the objective function in terms of the TPCs at the IoTNos and RC at the FC can be exploited using the BCD-based iterative framework, which designs the minimum-MSE TPC and combiner in an iterative fashion. \par
Employing the BCD algorithm, the optimization framework of the fully digital RC matrix assuming that all the TPC matrices are known reduces to the unconstrained problem of
 \begin{eqnarray}\label{Total}
 \underset{\mathbf{A}}{\text{minimize}}& \zeta\left(\mathbf{A}\right).
 \end{eqnarray}
Since the MSE objective in \eqref{MSE_2} is a quadratic convex function in terms of the RC matrix $\mathbf{A}$, the optimal RC matrix $\mathbf{A}^{*}$ can be obtained by differentiating it with respect to $\mathbf{A}$ and equating it to zero, which is given as
\begin{align}\label{combiner}
&\mathbf{A}^*=\mathbf{R}_{\theta}\left(\mathbf{C}^H_n\mathbf{P}^H_n\mathbf{H}^H_n\right)\Bigg[\sum_{n=1}^{N}\mathbf{H}_n\mathbf{P}_n\left(\mathbf{C}_n\mathbf{R}_{\theta}\mathbf{C}_n+\mathbf{R}_n\right)\nonumber \\
&\mathbf{P}^H_n\mathbf{H}^H_n+\sum_{n=1}^{N}\sum_{j=1,j\neq n}^{N}\mathbf{H}_n\mathbf{P}_n\mathbf{C}_n\mathbf{R}_{\theta}\mathbf{C}^H_j\mathbf{P}^H_j\mathbf{H}^H_j+\mathbf{R}_w\Bigg]^{-1}.
\end{align} 
Given the minimum-MSE fully digital RC, the TPC matrices $\mathbf{P}_n$ corresponding to each IoTNo $n$ can be designed by formulating the optimization problem of
\begin{align}\label{EQ:opt_pre}
&\underset{\left(\mathbf{P}_n\right)_{n=1}^N}{\text{minimize}} \quad {\zeta\left(\left(\mathbf{P}_n\right)_{n=1}^N\right)}
\nonumber\\
&\text{subject to } \text{Tr}\left[\mathbf{P}_{n}\left(\mathbf{C}_n\mathbf{R}_{\boldsymbol{\theta}}\mathbf{C}^H_n+\mathbf{R}_n\right)\mathbf{P}^H_{n}\right]\leq \rho_n,\;1\leq n \leq N.
\end{align}
The above optimization problem is convex in nature for a particular TPC matrix $\mathbf{P}_n$ corresponding to the $n$th IoTNo assuming that all the TPC matrices corresponding to the other IoTNos are known in addition to the RC $\mathbf{A}^*$. Upon invoking the Karush-Kuhn-Tucker (KKT) framework \cite{boyd2004convex} for the above problem, the Lagrangian function is given by
\begin{align}
&\mathcal{L}\left(\mathbf{P}_n,\lambda_n\right)=\text{Tr}\Bigg[\mathbf{A}\left(\sum_{n=1}^{N}\mathbf{H}_n\mathbf{P}_n\left(\mathbf{C}_n\mathbf{R}_{\theta}\mathbf{C}_n^H+\mathbf{R}_n\right)\mathbf{P}^H_n\mathbf{H}^H_n\right)\nonumber \\&\mathbf{A}^H+\mathbf{A}\left(\sum_{n=1}^{N}\sum_{j=1,j\neq n}^{N}\mathbf{H}_n\mathbf{P}_n\mathbf{C}_n\mathbf{R}_{\theta}\mathbf{C}^H_j\mathbf{P}^H_j\mathbf{H}^H_j\right)\mathbf{A}^H+\mathbf{A} \nonumber \\
&\mathbf{R}_w\mathbf{A}^H-\mathbf{A}\left(\sum_{n=1}^N \mathbf{H}_n\mathbf{P}_n \mathbf{C}_n\right)\mathbf{R}_{\theta}-\mathbf{R}_{\theta}\left(\sum_{n=1}^N\mathbf{C}^H_n\mathbf{P}^H_n\mathbf{H}^H_n\right)\nonumber \\
&\mathbf{A}^H+\mathbf{R}_{\theta}\Bigg]-\lambda_n\left[\text{Tr}\left[\mathbf{F}_{n}\left(\mathbf{C}_n\mathbf{R}_{\boldsymbol{\theta}}\mathbf{C}^H_n+\mathbf{R}_n\right)\mathbf{F}^H_{n}\right]- \rho_n\right],
\end{align}
where $\lambda_n$ is the dual variable corresponding to the $n$th sensor's power constraint. Upon employing the first order optimality KKT condition, the minimum-MSE TPC matrix corresponding to the $n$th IoTNo, which is denoted by $\mathbf{P}^*_n$, can be obtained as
\begin{align}\label{opt_pre}
&\mathbf{P}^*_n=\left[\mathbf{H}^H_n\mathbf{A}^H\mathbf{A}\mathbf{H}_n+\lambda_n\mathbf{I}\right]^{-1}\Bigg[\mathbf{H}^H_n\mathbf{A}^H\mathbf{R}_{\theta}\mathbf{C}^H_n-\nonumber \\
&\sum_{\substack{j=1\\j\neq n}}^N\mathbf{H}^H_n\mathbf{A}^H\mathbf{A}\mathbf{H}_j\mathbf{P}_j\left(\mathbf{C}_n\mathbf{R}_{\theta}\mathbf{C}_j^H\right)^H\Bigg]\left[\mathbf{C}_n\mathbf{R}_{\theta}\mathbf{C}^H_n+\mathbf{R}_n\right]^{-1}.
\end{align}
The optimal dual variable corresponding to each IoTNo $n$ can be obtained by solving the complementary slackness condition:
\begin{align}
\lambda_n\left[\text{Tr}\left[\mathbf{P}_{n}\left(\mathbf{C}_n\mathbf{R}_{\boldsymbol{\theta}}\mathbf{C}^H_n+\mathbf{R}_n\right)\mathbf{P}^H_{n}\right]- \rho_n\right]=0.
\end{align}
The calculation of dual variables is shown in Appendix \ref{appdx:dualv}.
%The above equation can be solved efficiently using the procedure described in \cite{1254038}. 
The above iterative algorithm terminates, when either a desired MSE level is achieved or a certain pre-set number of iterations have been performed. Once, the minimum-MSE fully-digital TPCs and RC matrices are derived, the next two subsections describe the SOMP algorithm based procedure of decomposing them into their RF and BB counterparts.
\subsection{SOMP based hybrid transceiver design}
\subsubsection{Precoder design}
This subsection describes the detailed procedure of decomposing the minimum-MSE fully-digital TPC matrix $\mathbf{P}_n$ corresponding to each IoTNo $n$ into its RF and BB counterparts $\mathbf{P}_{\text{RF},n}$ and $\mathbf{P}_{\text{BB},n}$, respectively. The corresponding optimization problem can be readily formulated as
\begin{align}
&\underset{\mathbf{P}_{\text{RF},n},\mathbf{P}_{\text{BB},n}}{\text{minimize}} \quad ||\mathbf{P}_{n}-\mathbf{P}_{\text{RF},n}\mathbf{P}_{\text{BB},n}||^2_F \nonumber\\
& \text{subject to }  
\quad \left \vert \mathbf{P}_{\text{RF},n}(i,j)\right\vert = \frac{1}{\sqrt{N_t}}\; \forall\; i,j.
\end{align}
The above problem is non-convex and intractable in nature, since the elements of the RF TPC must have constant magnitude. However, one can exploit an interesting relationship between the optimal TPC $\mathbf{P}_n$ and the transmit array response matrix $\mathbf{A}_{s,n}$, which is shown next in detail. It allows us to design $\mathbf{P}_{\text{RF},n}$ using the columns of the $\mathbf{A}_{s,n}$. To this end, using the expression for the minimum-MSE TPC $\mathbf{P}^*_n$ in \eqref{opt_pre} and subsequently substituting the SVD of the matrix $\mathbf{H}_n=\mathbf{U}_n\mathbf{\Sigma}_n\mathbf{V}^H_n$, one obtains
\begin{align}
&\mathbf{P}^*_n=\left[\mathbf{V}_n\mathbf{\Sigma}^H_n\mathbf{U}^H_n\mathbf{A}^H\mathbf{A}\mathbf{U}_n\mathbf{\Sigma}_n\mathbf{V}^H_n+\lambda_n\mathbf{I}\right]^{-1}\Bigg[\mathbf{V}_n\mathbf{\Sigma}^H_n\mathbf{U}^H_n \nonumber \\
&\mathbf{A}^H \mathbf{R}_{\theta}\mathbf{C}^H_n-\sum_{j=1,j\neq n}^N\mathbf{V}_n\mathbf{\Sigma}^H_n\mathbf{U}^H_n\mathbf{A}^H\mathbf{A}\mathbf{U}_j\mathbf{\Sigma}_j\mathbf{V}^H_j\mathbf{P}_j\big(\mathbf{C}_n\mathbf{R}_{\theta}\nonumber \\
&\mathbf{C}_j\big)^H\Bigg]\left[\mathbf{C}_n\mathbf{R}_{\theta}\mathbf{C}^H_n+\mathbf{R}_n\right]^{-1} \nonumber
\end{align}
\begin{align}
&=\mathbf{V}_n\left[\mathbf{\Sigma}^H_n\mathbf{U}^H_n\mathbf{A}^H\mathbf{A}\mathbf{U}_n\mathbf{\Sigma}_n+\lambda_n\mathbf{I}\right]^{-1}\mathbf{V}^H_n\Bigg[\mathbf{V}_n\mathbf{\Sigma}^H_n\mathbf{U}^H_n\mathbf{A}^H \nonumber \\
&\mathbf{R}_{\theta}\mathbf{C}^H_n-\sum_{j=1,j\neq n}^N\mathbf{V}_n\mathbf{\Sigma}^H_n\mathbf{U}^H_n\mathbf{A}^H\mathbf{A}\mathbf{U}_j\mathbf{\Sigma}_j\mathbf{V}^H_j\mathbf{P}_j\big(\mathbf{C}_n\mathbf{R}_{\theta}\nonumber \\
&\mathbf{C}_j^H\big)^H\Bigg]\left[\mathbf{C}_n\mathbf{R}_{\theta}\mathbf{C}^H_n+\mathbf{R}_n\right]^{-1}.
\end{align}
This implies that the column space of the TPC matrix $\mathbf{P}^*_n$ corresponding to the $n$th IoTNo lies in the column space of the unitary matrix $\mathbf{V}_n$. Moreover, the columns of $\mathbf{V}_n$ span the row space of the channel matrix $\mathbf{H}_n$. Thus, it can be concluded that 
\begin{align}
\mathcal{C}\left(\mathbf{P}_n\right) \subseteq \mathcal{R}\left(\mathbf{H}_n\right).
\end{align}
It can be observed from \eqref{Channel:EQ} that the row-space of the mmWave MIMO channel matrix $\mathbf{H}_n$ is a subset of the column-space of the transmit array response matrix $\mathbf{A}_{s,n}$. Thus, we can write 
\begin{align}
\mathcal{C}\left(\mathbf{P}_n\right) \subseteq \mathcal{R}\left(\mathbf{H}_n\right) \subseteq \mathcal{C}\left(\mathbf{A}_{s,n}\right).
\end{align}
Hence, we can design the RF TPC matrix $\mathbf{P}_{\text{RF},n}$ by extracting the $N_{\text{RF}}^n$ columns of $\mathbf{A}_{s,n}$, which by default satisfies the constant-magnitude condition. Thus, the optimization problem of designing the BB TPC $\mathbf{P}_{\text{BB},n}$ can be formulated as
\begin{equation}\label{SOMP_Precoder}
\begin{aligned} 
\underset{{\bar{\mathbf{P}}_{\text{BB},n}}}{\text{minimize}} \quad & ||\mathbf{P}_{n}-\mathbf{A}_{s,n}\bar{\mathbf{P}}_{\text{BB},n}||^2_F \\
\textrm{subject to} \quad & 
\left \vert \left \vert \text{diag}\left(\bar{\mathbf{P}}_{\text{BB},n}\bar{\mathbf{P}}^H_{\text{BB},n}\right) \right\vert \right\vert_0 = N_{\text{RF}}^n, \\
%&||\mathbf{P}_{\text{RF},n}\bar{\mathbf{P}}_{\text{BB},n}\mathbf{D}_n||^2_F \leq \rho_n,
\end{aligned}
\end{equation}
where the first constraint implies that the matrix $\bar{\mathbf{P}}_{\text{BB},n}$ is block-sparse in nature, since only $N_{\text{RF}}$ rows out of $N$ rows are non-zero. Hence, similar to Ayach et al. \cite{el2014spatially}, the simultaneous orthogonal matching pursuit (SOMP) algorithm is employed for design of the RF and BB
components of the TPC, since the latter is a simultaneously sparse matrix.
% Furthermore, the second constraint ensures that the resultant TPC matrix satisfies the power constraint of the $n$th IoTNo.
Algorithm-1 describes the proposed SOMP-based hybrid TPC design technique in detail. The next subsection describes the detailed procedure to decompose the minimum-MSE RC matrix $\mathbf{A}$ into its RF and BB counterparts.
\begin{algorithm}[t]
\caption{ SOMP- based hybrid precoder design}\label{alg:algorithm1}
\begin{algorithmic}[1]
\Require{$\left\lbrace \mathbf{P}_n^{\mathrm{opt}}\right\rbrace$ and $N_{\text{RF}}^n$ }\; $\forall$ $n$
\For{$1 \leq n \leq N$}
\State $\mathbf{P}_{\text{RF},n}=[\  ]$
\State $\mathbf{P}_{\text{res}}=\mathbf{P}_n$
\For{$k \leq N_{\text{RF}}^n$}
   \State$\boldsymbol{\Psi}=\mathbf{A}_{s,n}^H\mathbf{P}_{\text{res}}$
   \State $m={\textrm{arg max}}_{m=1, ..., N_{cl}}(\boldsymbol{\Psi}\boldsymbol{\Psi}^H)_{l,l}$ 
   \State $\mathbf{P}_{\text{RF},n}=[\mathbf{P}_{\text{RF},n}\mid \mathbf{A}_{s,n}^{(n)}]$
   \State $\mathbf{P}_{\text{BB},n}=(\mathbf{P}_{\text{RF},n}^H\mathbf{P}_{\text{RF},n})^{-1}\mathbf{P}_{\text{RF},n}^H\mathbf{P}_n$
   \State $\mathbf{P}_{\text{res}}=\frac{\mathbf{P}_n-\mathbf{P}_{\text{RF},n}\mathbf{P}_{\text{BB},n}}{\|\mathbf{P}_n-\mathbf{P}_{\text{RF},n}\mathbf{P}_{\text{BB},n}\|_F}$
\EndFor
\State $\mathbf{P}_{{\text{BB}},n}=\frac{\mathbf{P}_{{\text{BB}},n}\left\lvert\left\lvert\mathbf{P}_{n}\right\rvert\right\rvert_F}{\left\lvert\left\lvert\mathbf{P}_{{\text{RF}},n}\mathbf{P}_{{\text{BB}},n}\right\rvert\right\rvert_F}$
\State \Return{$\mathbf{P}_{\text{RF},n}$ , $\mathbf{P}_{\text{BB},n}$}
\State $\mathbf{P}_{\text{RF}}=\text{blkdiag}\left(\mathbf{P}_{\text{RF}},\mathbf{P}_{\text{RF},n}\right)$
\State $\mathbf{P}_{\text{BB}}=\text{blkdiag}\left(\mathbf{P}_{\text{BB}},\mathbf{P}_{\text{BB},n}\right)$
\EndFor
\end{algorithmic}
\end{algorithm}
\subsubsection{Combiner design}
Once the optimal RF and baseband TPC matrices corresponding to each IoTNo $n$ are derived, the next objective is to decompose the minimum-MSE RC matrix into its RF and baseband counterparts. The corresponding optimization is formulated as
\begin{equation}
\begin{aligned} 
\underset{{\mathbf{A}_{\text{RF}}},{\mathbf{A}_{\text{BB}}}}{\text{minimize}} \quad & ||\boldsymbol{\theta}-{\mathbf{A}^H_{\text{BB}}},{\mathbf{A}^H_{\text{RF}}}\mathbf{y}||^2_F \\
\textrm{subject to} \quad & 
\left \vert \mathbf{A}_{\text{RF}}(i,j)\right\vert = \frac{1}{\sqrt{N_r}}\; \forall\; i,j.
\end{aligned}
\end{equation}
Upon, ignoring the constraint in the above optimization problem, the optimal solution of the unconstrained optimization problem is the RC matrix given in \eqref{combiner}. As shown in \cite{el2014spatially}, the above optimization problem may also be modified as
\begin{equation}
\begin{aligned} 
\underset{{\mathbf{A}_{\text{RF}}},{\mathbf{A}_{\text{BB}}}}{\text{minimize}} \quad & ||\mathbf{R}^{\frac{1}{2}}_{yy}\left(\mathbf{A}-{\mathbf{A}_{\text{RF}}}{\mathbf{A}_{\text{BB}}}\right)||^2_F \\
\textrm{subject to} \quad & 
\left \vert \mathbf{A}_{\text{RF}}(i,j)\right\vert = \frac{1}{\sqrt{N_r}}\; \forall\; i,j.
\end{aligned}
\end{equation}
The above problem is non-convex due to the constant-magnitude constraint. However, similar to the previous subsection, it can be shown that the column space of the RC matrix $\mathbf{A}$ lies in the column space of the concatenated mmWave MIMO channel $\mathbf{H}$. The detailed justification for this follows next. The optimal fully digital RC expression given in \eqref{combiner} can also be written equivalently as
\begin{align}
\mathbf{A}^H=\left[\mathbf{HP}\left(\mathbf{C}\mathbf{R}_{\theta}\mathbf{C}^H+\mathbf{R}_{n}\right)\mathbf{P}^H\mathbf{H}^H+\sigma^2_n\mathbf{I}\right]^{-1}\mathbf{HPC}\mathbf{R}_{\theta},
\end{align}
where the block-diagonal TPC matrix $\mathbf{P} \in \mathbb{C}^{NN_t \times Nl}$ is defined as
\begin{align*}
\mathbf{P}&=\text{blkdiag}\left[\mathbf{P}_1, \mathbf{P}_2,\hdots,\mathbf{P}_N\right], 
\end{align*}
and $\mathbf{P}_n=\mathbf{P}_{\text{RF},n}\mathbf{P}_{\text{BB},n}$. Substituting $\mathbf{R}_n=\sigma^2_n\mathbf{I}_l$, and using the property $\left(\mathbf{I}+\mathbf{GQ}\right)^{-1}\mathbf{G}=\mathbf{G}\left(\mathbf{I}+\mathbf{QG}\right)^{-1}$, for any suitable matrices $\mathbf{G}$ and $\mathbf{Q}$, the optimal fully-digital RC expression above can further be recast as
\begin{align}
\mathbf{A}^H=\mathbf{HP}\left[\left(\mathbf{C}\mathbf{R}_{\theta}\mathbf{C}^H+\mathbf{R}_{n}\right)\mathbf{P}^H\mathbf{H}^H\mathbf{HP}+\sigma^2_n\mathbf{I}\right]^{-1}\mathbf{C}\mathbf{R}_{\theta},
\end{align}
where $\mathbf{G}=\mathbf{HP}$ and $\mathbf{Q}=\left(\mathbf{C}\mathbf{R}_{\theta}\mathbf{C}^H+\mathbf{R}_{n}\right)\mathbf{P}^H\mathbf{H}^H$. It can now be observed that any column of $\mathbf{A}^H$ can be written as a linear combination of the columns of $\mathbf{H}$. Furthermore, the columns of the matrix $\mathbf{H}$ are linear combinations of the columns of the receive array response matrix $\mathbf{A}_{\text{FC}}$. Thus, it can be concluded that 
\begin{align}
\mathcal{C}\left(\mathbf{A}^H\right) \subseteq \mathcal{C}\left(\mathbf{H}\right) \subseteq \mathcal{C}\left(\mathbf{A}_{\text{FC}}\right).
\end{align}
Hence, we can design $\mathbf{A}_{\text{RF}}$ by extracting $N_{\text{RF}}$ columns from $\mathbf{A}_{\text{FC}}$, which also satisfy the constant gain condition on the elements of the RF RC matrix. Subsequently, one can develop the BB RC design problem as
\begin{equation}\label{EqSOMPC}
\begin{aligned} 
\underset{{\mathbf{A}_{\text{BB}}}}{\text{minimize}} \quad & ||\mathbf{R}^{\frac{1}{2}}_{yy}\left(\mathbf{A}-\mathbf{A}_{\text{FC}}\bar{\mathbf{A}}_{\text{BB}}\right)||^2_F \\
\textrm{subject to} \quad & 
\left \vert \left \vert \text{diag}\left(\bar{\mathbf{A}}_{\text{BB}}\bar{\mathbf{A}}^H_{\text{BB}}\right) \right\vert \right\vert_0 = N_{\text{RF}}.
\end{aligned}
\end{equation}
The above optimization problem is once again a sparse signal recovery problem and can be solved similar to \eqref{SOMP_Precoder} upon employing the SOMP algorithm. The complete iterative procedure of designing the hybrid transceiver is summarized in Algorithm-2. The next subsection presents a proof of the convergence of the proposed BCD-based iterative algorithm.

	\subsection{Convergence analysis}\label{convergence}
The proposed BCD-based iterative transceiver scheme's theoretical convergence can be proved as follows. Due to the fact that the optimization problems of the fully-digital RC and TPC matrices in \eqref{Total} and \eqref{EQ:opt_pre} are convex in nature, the following inequalities hold:
\begin{align}
&\mathrm{MSE}\Big(\mathbf{A}(i),\{\mathbf{P}_n(i)\}_{n=1}^N\Big) \geq 
\underset{\mathbf{A}}{\text{min}}\; \mathrm{MSE}\Big(\mathbf{A}|\{\mathbf{P}_n(i)\}_{n=1}^N\Big)  \nonumber \\
&=\mathrm{MSE}\Big(\mathbf{A}(i+1),\{\mathbf{P}_n(i)\}_{n=1}^N\Big) \nonumber \\ &\geq 
\underset{\mathbf{P}_n}{\text{min}}\; \mathrm{MSE}\Big(\mathbf{P}_n|\mathbf{A}(i+1),\{\mathbf{P}_k(i+1)\}_{k=1}^{N-1}, \{\mathbf{P}_k(i)\}_{k=n+1}^N\Big) \nonumber \\
&=\mathrm{MSE}\Big(\mathbf{P}_n(i+1),\mathbf{A}(i+1)\Big).
\end{align}
Hence, $\mathrm{MSE}\Big(\{\mathbf{A}\}^{(i)},\{\mathbf{P}_n\}^{(i)}\Big)$ is a monotonically decreasing sequence and it is lower-bounded by zero. This proves that the algorithm converges. On convergence, the SOMP algorithm is employed for obtaining the RF and BB TPCs and RCs from their respective fully-digital counterparts. 
\begin{algorithm} [!t]\label{Algo}
		\caption{BCD-based iterative algorithm for hybrid transceiver design}		
		\begin{algorithmic}[1]
	\State \textbf{Input}: Observation vector $\mathbf{y}$, maximum iterations $i_{\text{max}}=40$ and desired accuracy $\epsilon=0.0001$.
		\State \textbf{Initialization}: $i=1$, initialize precoding matrices $\left\{\mathbf{P}_n(i-1)\right\}_{n=1}^{N}$ randomly.
		    \While {$\parallel\widehat{\boldsymbol{\theta}}^{(n)}-{\boldsymbol{\theta}}^{(n)}\parallel_{2}\geq \epsilon$ and $i < i_{max}$} 
		    
		    \State Evaluate the fully digital combiner matrix $\mathbf{A}(i)$ using \eqref{combiner}.
			\State Evaluate the optimal precoding matrices $\left\{\mathbf{P}_n(i)\right\}_{n=1}^{N}$ using \eqref{opt_pre}.
\State update $i=i+1$.
			\EndWhile
			\State Employ the SOMP algorithm detailed in Subsection-IV-A to decompose the fully digital combiner/ precoder into the corresponding hybrid combiner/ precoder.
			\State \textbf{Output}: $\mathbf{A}_{\text{RF}}$, $\mathbf{A}_{\text{BB}}$, $\left\{\mathbf{P}_{\text{RF},n}\right\}_{n=1}^{N}$, $\left\{\mathbf{P}_{\text{BB},n}\right\}_{n=1}^{N}$.
		\end{algorithmic}
	\end{algorithm}
	The
next section develops robust transceiver designs for a scenario associated with CSI uncertainty. 
%The next section extends the transceiver design developed in this section to a practical scenario where the CSI between each sensor and the FC is not known perfectly. The CSI uncertainty is modeled using the popular stochastic CSI uncertainty model. Subsequently, the robust transceiver is designed to minimize the average MSE at the FC subject to individual sensor power constraints.
\section{Robust Hybrid Transceiver Design for a Noisy FC}\label{Sec-V}
In practical scenarios, it is impossible to have perfect CSI knowledge between each IoTNo and the FC due to the limited pilot overhead and quantization errors. Hence, in a practical system, one has to take the CSI uncertainty into account. Therefore, motivated by \cite{8606437,9079551,9448465}, we model the channel between each IoTNo and the FC as 
\begin{align}\label{EQ:H}
\mathbf{H}_n=\widehat{\mathbf{H}}_n+\Delta{\mathbf{H}}_n,
\end{align}
where $\widehat{\mathbf{H}}_n$ denotes the available channel estimate and $\Delta{\mathbf{H}}_n$ represents the estimation error matrix whose elements obey the distribution $\mathcal{CN}\left(0,\sigma_H^2\right)$.
%The result given in the following lemma will be used in this section.
%\begin{lem}
%For a matrix $\mathbf{X} \in \mathbb{C}^{r \times t}$, which can also be represented as  $\mathbf{X}=\widehat{\mathbf{X}}+\Delta{\mathbf{X}}$, where the matrix $\widehat{\mathbf{X}}$ is already known while the elements of the matrix 
%$\Delta{\mathbf{X}}$ follows the distribution $\mathcal{CN}(0,\sigma^2)$, then it follows that 
%\begin{align}\label{Lemma}
%\mathbb{E}\left[\mathbf{X}\mathbf{Z}\mathbf{Z}^H\mathbf{X}^H\right]=\widehat{\mathbf{X}}\mathbf{Z}\mathbf{Z}^H\widehat{\mathbf{X}}^H+\sigma^2_H\mathrm{Tr}\left[\mathbf{Z}\mathbf{Z}^H\right]\mathbf{I}_r.
%\end{align}
%\end{lem} 
\begin{figure*}
\centering
\subfloat[]{\includegraphics[width = 0.45\linewidth]{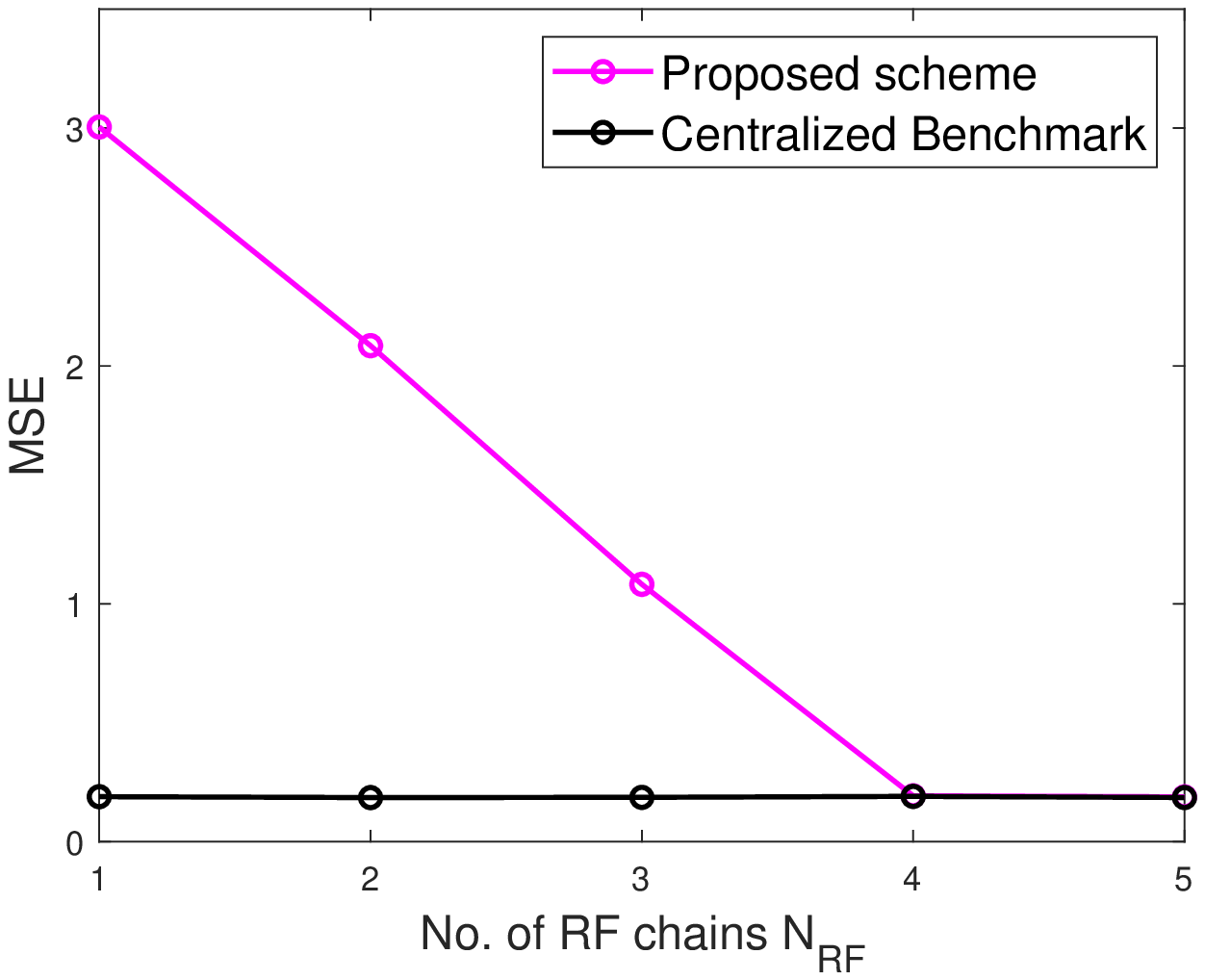}}
\hfil
\hspace{10pt}\subfloat[]{\includegraphics[width = 0.45\linewidth]{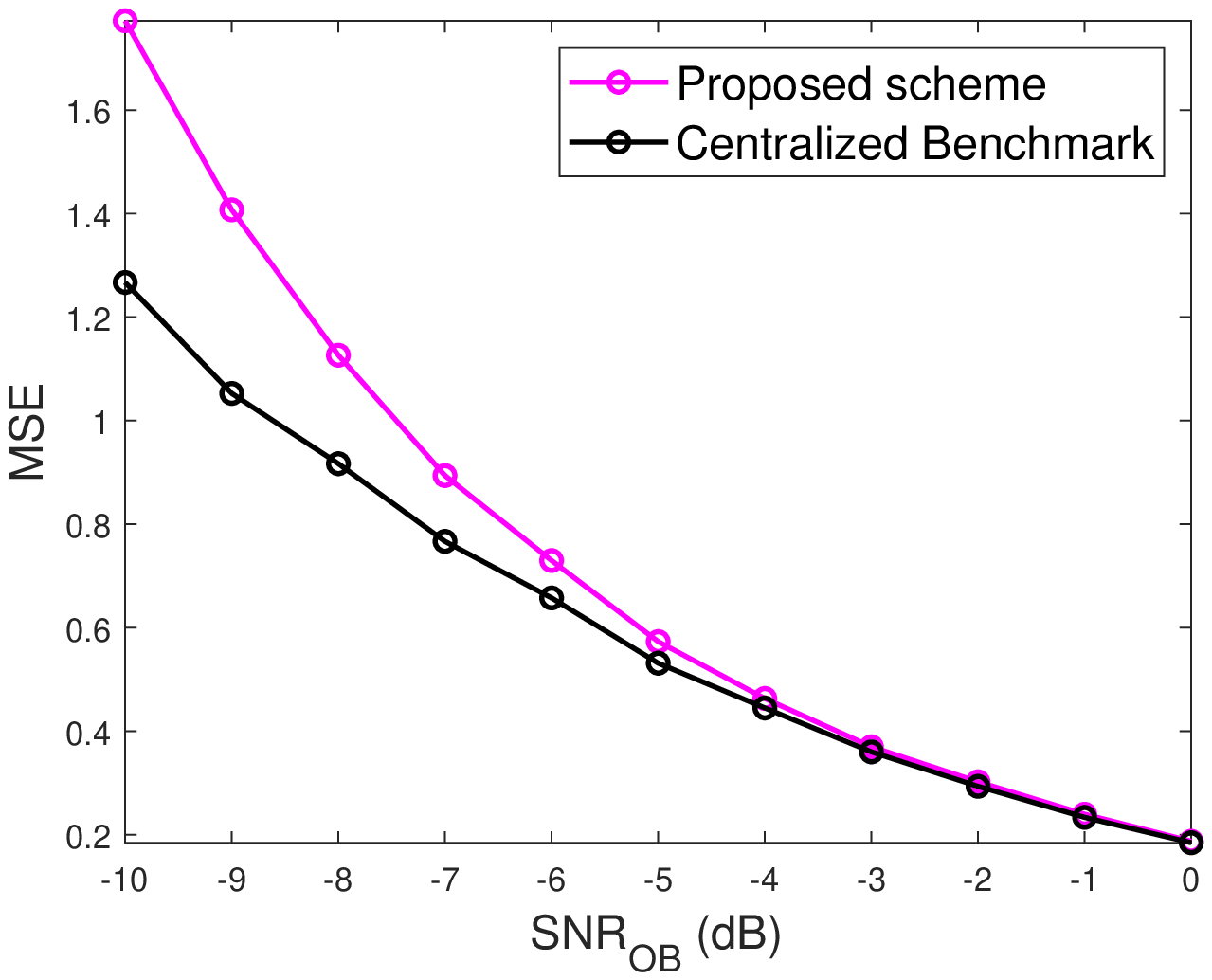}}
\caption{$\left(a\right) $ MSE versus number of RF chains $N_{\text{RF}}^n$ at $\mathrm{SNR}_{\text{OB}}=0$ dB $\left(b\right)$ MSE versus $\mathrm{SNR}_{\text{OB}}$ when $N_{\text{RF}}^n=4$, for the mmWave MIMO IoTNe with $N=25, N_{t} = 5, N_{r} = 16, q=4, l =2, K=5$.}
\label{fig:NMSE:static}
\end{figure*}
Substituting $\mathbf{H}_n$ from \eqref{EQ:H} into the expression in \eqref{MSE_2}, the expression for the resultant average MSE defined as $\bar{\zeta}\left(\mathbf{A},\mathbf{P}_n\right)=\mathbb{E}_{\Delta{\mathbf{H}}_n}\zeta\left(\mathbf{A},\mathbf{P}_n\right)$, is given by
\begin{align}\label{MSE_Robust}
&\bar{\zeta}\left(\mathbf{A},\mathbf{P}_n\right)=\text{Tr}\Bigg[\mathbf{A}\left(\sum_{n=1}^{N}\widehat{\mathbf{H}}_n\mathbf{P}_n\left(\mathbf{C}_n\mathbf{R}_{\theta}\mathbf{C}_n^H+\mathbf{R}_n\right)\mathbf{F}^H_n\widehat{\mathbf{H}}^H_n\right) \nonumber \\& \mathbf{A}^H + \mathbf{A}\left(\sum_{n=1}^{N}\sigma^2_H \text{Tr}\left[\mathbf{P}_n\left(\mathbf{C}_n\mathbf{R}_{\theta}\mathbf{C}_n^H+\mathbf{R}_n\right)\mathbf{P}^H_n\right]\right)\mathbf{A}^H + \mathbf{R}_{\theta}\nonumber \\& +\mathbf{A}\mathbf{R}_w\mathbf{A}^H +\mathbf{A}\left(\sum_{n=1}^{N}\sum_{\substack{j=1\\j\neq n}}^{N}\widehat{\mathbf{H}}_n\mathbf{P}_n\mathbf{C}_n\mathbf{R}_{\theta}\mathbf{C}^H_j\mathbf{P}^H_j\widehat{\mathbf{H}}^H_j\right)\mathbf{A}^H \nonumber \\
&-\mathbf{A}\left(\sum_{n=1}^N \widehat{\mathbf{H}}_n\mathbf{P}_n \mathbf{C}_n\right)\mathbf{R}_{\theta}-\mathbf{R}_{\theta}\left(\sum_{n=1}^N\mathbf{C}^H_n\mathbf{P}^H_n\widehat{\mathbf{H}}^H_n\right)\mathbf{A}^H\Bigg].
\end{align}
The detailed steps are shown in Appendix \ref{appdx:MSE_Robust}. Hence, the pertinent optimization problem for designing the hybrid transceiver in this framework associated with CSI uncertainty so that the average MSE is minimized at the FC is similar to \eqref{EQ:opt_NFC}, with the objective function ${\zeta}\left(\mathbf{A},\mathbf{P}_n\right)$ replaced by $\bar{\zeta}\left(\mathbf{A},\mathbf{P}_n\right)$. The BCD-based iterative framework can once again be employed to design the fully-digital robust transceiver followed by the SOMP algorithm to yield the robust hybrid TPC and RC matrices. One can determine the robust RC matrix at the FC for a given set of robust TPC matrices $\{\mathbf{P}_n\}_{n=1}^{N}$, by setting $\frac{\partial  \bar{\zeta}\left(\mathbf{A}\right)}{\partial \mathbf{A}}=\mathbf{0} $, as
\begin{align}\label{combiner_robust}
&\mathbf{A}^*=\mathbf{R}_{\theta}\left(\mathbf{C}^H_n\mathbf{P}^H_n\widehat{\mathbf{H}}^H_n\right)\Bigg[\sum_{n=1}^{N}\widehat{\mathbf{H}}_n\mathbf{P}_n\left(\mathbf{C}_n\mathbf{R}_{\theta}\mathbf{C}_n^H+\mathbf{R}_n\right)\nonumber \\
&\mathbf{P}^H_n\widehat{\mathbf{H}}^H_n+\sum_{n=1}^{N}\sigma^2_H\text{Tr}\left[\mathbf{P}_n\left(\mathbf{C}_n\mathbf{R}_{\theta}\mathbf{C}_n^H+\mathbf{R}_n\right)\mathbf{P}^H_n\right]\nonumber \\&+\sum_{n=1}^{N}\sum_{j=1,j\neq n}^{N}\widehat{\mathbf{H}}_n\mathbf{P}_n\mathbf{C}_n\mathbf{R}_{\theta}\mathbf{C}^H_j\mathbf{P}^H_j\widehat{\mathbf{H}}^H_j+\mathbf{R}_w\Bigg]^{-1}.
\end{align}
The optimization problem of designing the robust TPC matrices, given the fully-digital RC matrix, is equivalent to \eqref{EQ:opt_pre} with ${\zeta\left(\{\mathbf{P}_n\}_{n=1}^{N}\right)}$ replaced by $\bar{\zeta}\left(\{\mathbf{P}_n\}_{n=1}^{N}\right)$. This can be solved along similar lines by invoking the KKT framework of \cite{boyd2004convex} to obtain the robust fully-digital TPC matrices, given as
\begin{align}\label{opt_pre_rbst}
\mathbf{P}^*_n=&\left[\widehat{\mathbf{H}}^H_n\mathbf{A}^H\mathbf{A}\widehat{\mathbf{H}}_n+\sigma^2_H\text{Tr}\left[\mathbf{A}^H\mathbf{A}\right]+\lambda_n\mathbf{I}\right]^{-1}\Bigg[\widehat{\mathbf{H}}^H_n\mathbf{A}^H\mathbf{R}_{\theta}\nonumber \\
&\mathbf{C}^H_n-\sum_{j=1,j\neq n}^L\widehat{\mathbf{H}}^H_n\mathbf{A}^H\mathbf{A}\widehat{\mathbf{H}}_j\mathbf{P}_j\left(\mathbf{C}_n\mathbf{R}_{\theta}\mathbf{C}_j^H\right)^H\Bigg]\bigg[\mathbf{C}_n \nonumber \\&\mathbf{R}_{\theta}\mathbf{C}^H_n+\mathbf{R}_n\bigg]^{-1}.
\end{align}
As described in the previous section, the SOMP-based algorithm can once again be employed for deriving the hybrid TPC and RC matrices for this scenario in the face of CSI uncertainty. The next subsection derives the centralized MMSE benchmark, which acts as a lower bound for the LDE schemes.

\subsection{Centralized MMSE benchmark}
The
centralized MMSE benchmark represents the best achievable
performance where all the IoTNo observations are directly
available at the FC.
%The best estimation performance can be achieved when the observation $\mathbf{x}_n$ corresponding to each sensor $n$ is available directly at the FC without any distortion.
The observation vector $\mathbf{x} \in \mathbb{C}^{Nl \times 1}$ available at the FC can be modeled as 
\begin{align}
\mathbf{x}=\mathbf{C}\boldsymbol{\theta}+\mathbf{v},
\end{align} 
where $\mathbf{x}=[\mathbf{x}^T_1,\mathbf{x}^T_2,\hdots,\mathbf{x}^T_L]^T \in \mathbb{C}^{Nl \times 1}$. 
%The MMSE estimate evaluated at the FC for this system yields the best possible $\mathrm{MSE}$, which is expressed as
The MSE is formulated as
\begin{align}
\mathrm{MSE}_{\text{MMSE}}=\mathrm{Tr}\left[\left(\mathbf{I}_{q}+\mathbf{C}^H\mathbf{C}\right)^{-1}\right].\label{eq:centr_bench}
\end{align}
The MSE bound above can be used as a benchmark for the evaluation of the performance of the proposed LDE schemes. The next section discusses the simulation results to illustrate the effectiveness of the various designs proposed.
\section{Simulation Results} \label{Sec-VI} 
 \begin{figure*}
\centering
\subfloat[]{\includegraphics[width=0.45\linewidth]{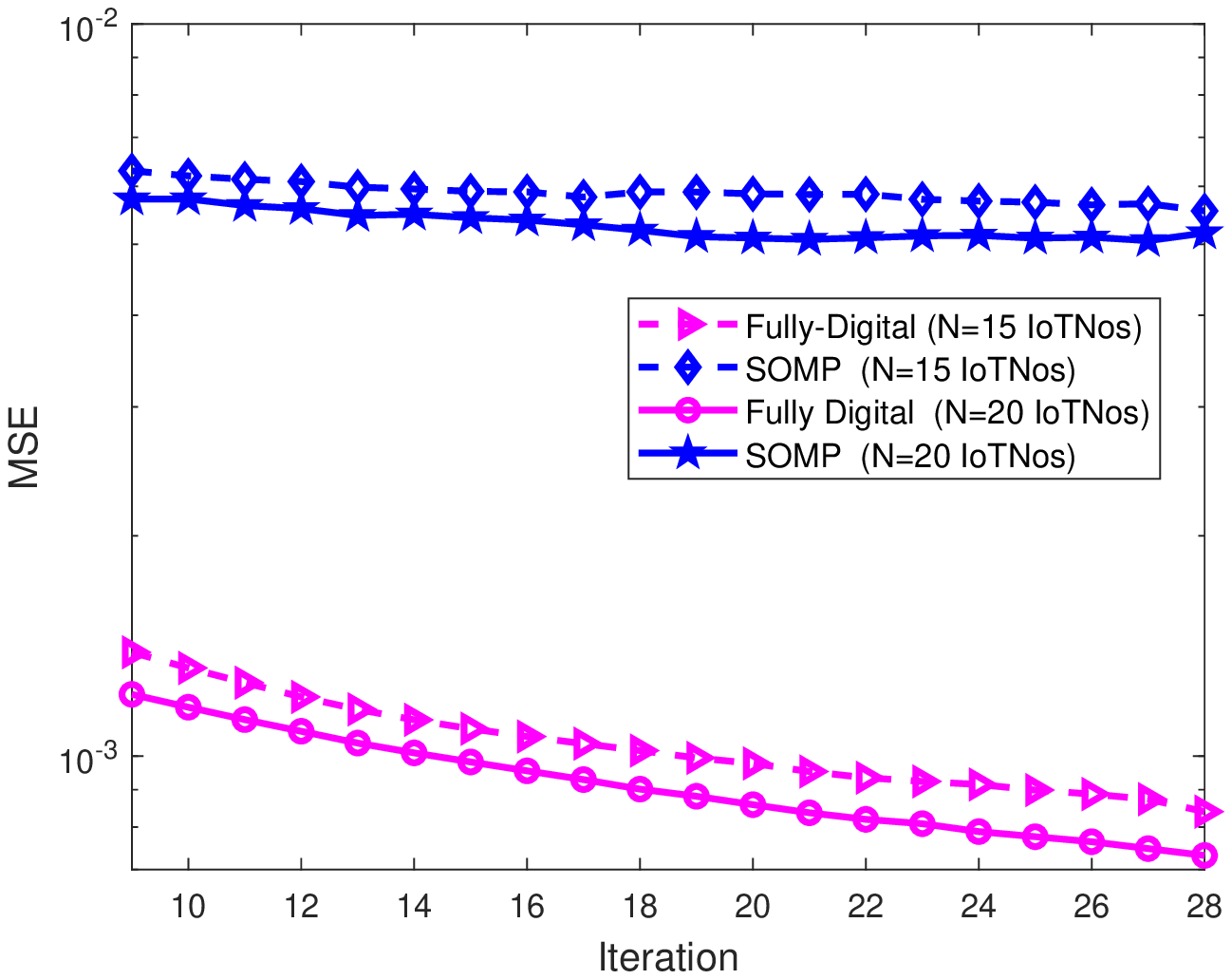}}
\hfil
\hspace{10pt}\subfloat[]{\includegraphics[width=0.45\linewidth]{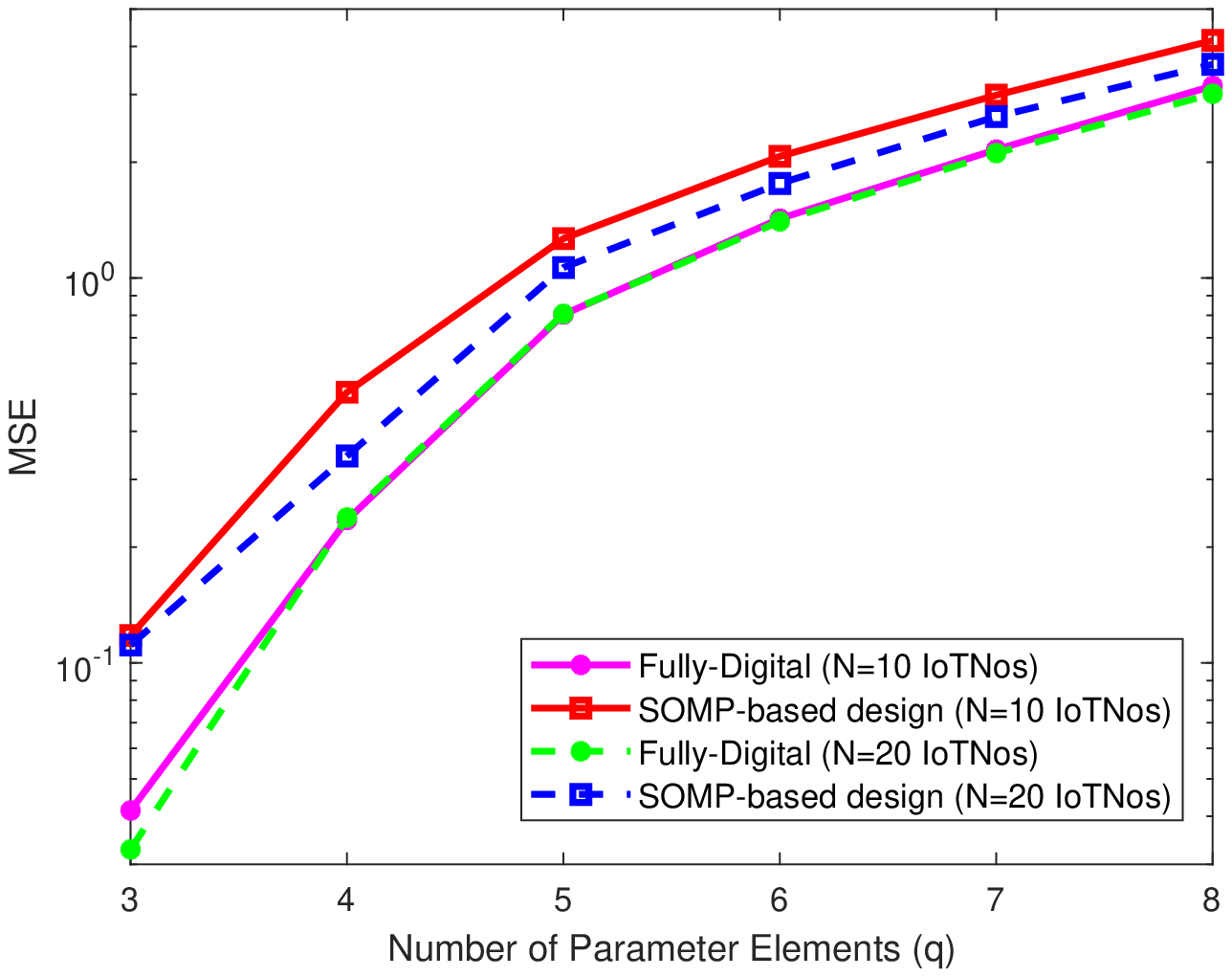}}
\caption{$\left(a\right) $ MSE versus the number of BCD iterations for different number of IoTNos $N \in \{10,20\}$ $\left(b\right)$ MSE versus the number of elements in the parameter vector $\boldsymbol{\theta}$ for different number of IoTNos $N \in \{10,20\}$ $\left(b\right)$.}
\label{fig:4}
\end{figure*}
A mmWave MIMO IoTNe is considered where each IoTNo is equipped with $N_t=5$ TAs, and the FC has $N_r= 10$ RAs.
%The inter-antenna spacings of the antenna arrays both at the sensors and the FC are fixed as $d_{T} = d_{R} = \lambda/2$. 
The number of clusters in the mmWave MIMO channel is set to $K=5$ clusters. The path-gains $\alpha_{k,n}$ are randomly generated from the distribution $\mathcal{CN}(0,1)$. The dimension of the parameter vector to be estimated is set to $q=3$ and the number of observations made by each IoTNo is set to $l=2$. The observation SNR is defined as $\mathrm{SNR}_{\text{OB}}=\frac{1}{\sigma_l^2}$ whereas the $\mathrm{SNR}_{\text{FC}}=\frac{1}{\sigma_n^2}$. The elements of the observation matrix $\mathbf{C}_n$ have been randomly generated from the distribution $\mathcal{CN}(0,1)$.  \par
 \begin{figure*}
\centering
\subfloat[]{\includegraphics[width=0.45\linewidth]{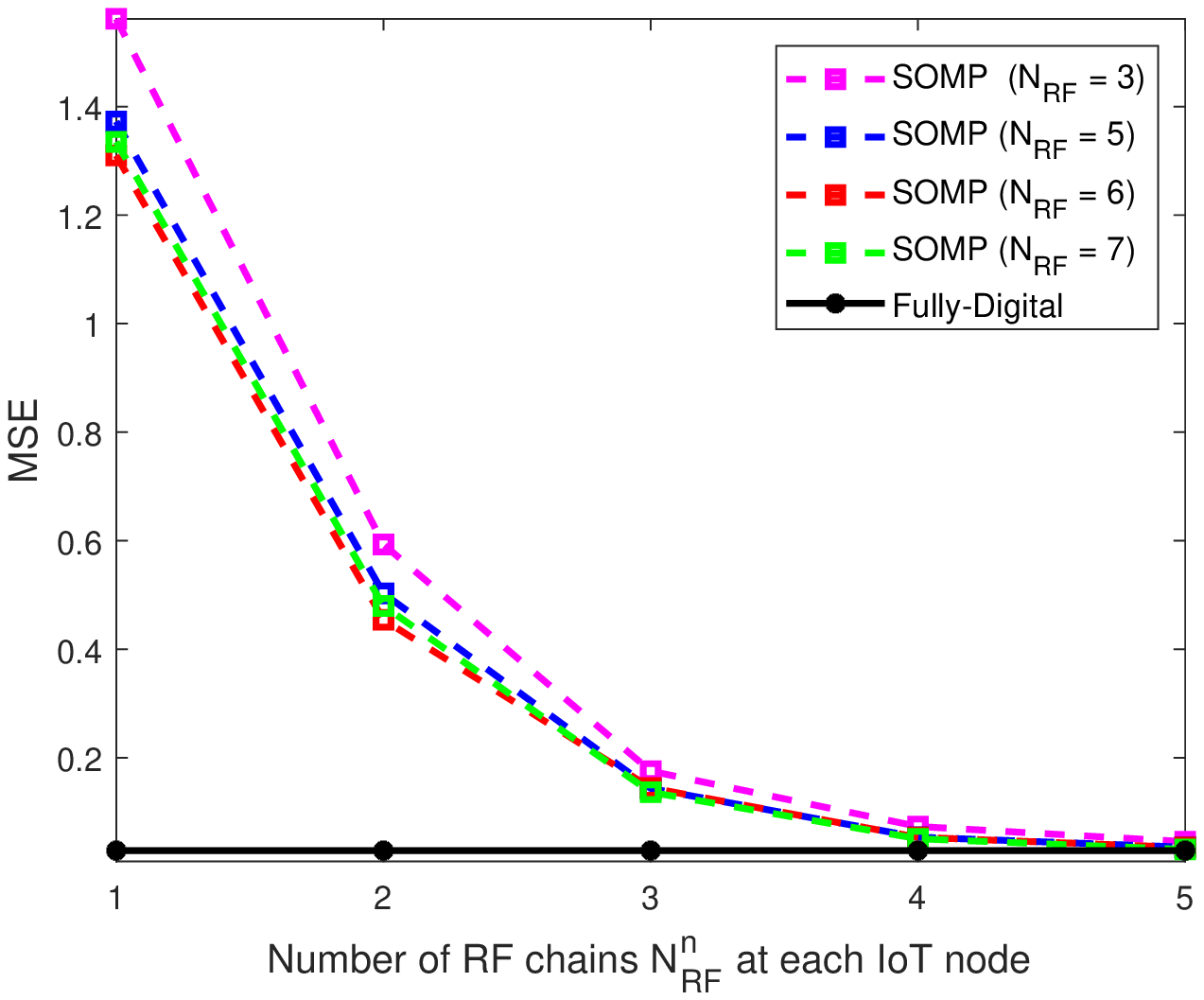}}
\hfil
\hspace{10pt}\subfloat[]{\includegraphics[width=0.45\linewidth]{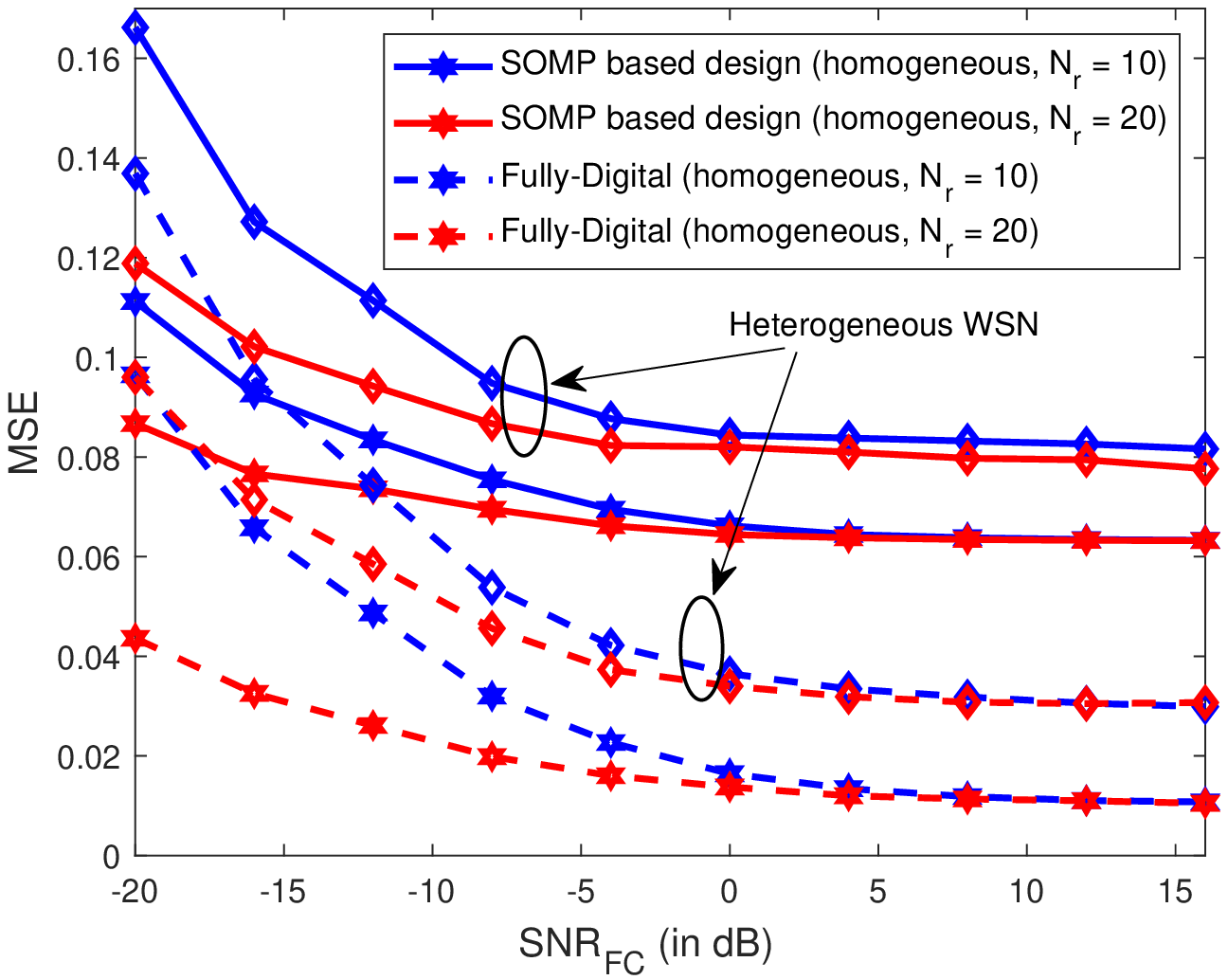}}
\caption{$\left(a\right) $ MSE versus the number of RF chains ($N_{\text{RF}}^n$) for $N=20$ IoTNos $\left(b\right)$ MSE versus $\text{SNR}_{\text{FC}}$ for different number of RAs $N_r \in \left\lbrace10,20\right\rbrace$.}
\label{fig:5}
\end{figure*} 
 Fig. \ref{fig:NMSE:static}(a) illustrates the MSE performance of the proposed hybrid transceiver against the number of RF chains $N_{\text{RF}}$ employed at each IoTNo at $\mathrm{SNR}_{\text{OB}}=0$ dB. Observe that the MSE performance improves upon increasing the number of RF chains, and as claimed in \eqref{eq:6} it matches the centralized MMSE benchmark derived in \eqref{eq:centr_bench} when the number of RF chains $N_{\text{RF}}$ becomes equal to the parameter length $q$. It can also be readily observed that increasing the number of RF chains beyond the parameter dimension does not result in any improvement in the MSE performance at the FC, thus validating our analytical result. \par
 Fig. \ref{fig:NMSE:static}(b) illustrates the MSE versus $\mathrm{SNR}_{\text{OB}}$, when the number of RF chains $N_{\text{RF}}$ is set equal to the length of the parameter $q$. It can be observed from the figure that the MSE of both the proposed scheme and the centralized benchmark improve upon increasing the observation SNR, which is along expected lines. Moreover, the gap between the proposed scheme and the centralized benchmark decreases upon increasing the observation SNR and matches the centralized benchmark at high SNRs, thus validating the effectiveness of the proposed scheme.\par
 Fig. \ref{fig:4}(a) depicts the MSE performance as a function of the number of the BCD-iterations for the design proposed in Section-IV for a noisy FC with perfect CSI. The number of IoTNos is $N \in \{15,20\}$ for this particular plot. To benchmark the SOMP-based hybrid transceiver design's performance, its corresponding fully-digital performance has also been shown. The MSE performance improves as the number of iterations increases. Furthermore, the MSE also decreases as the number of sensors $K$ increases, a trend that is along expected lines due to the availability of a larger number of observations.\par
Fig. \ref{fig:4}(b) illustrates the MSE performance against the dimension of the underlying parameter vector of interest for different number of IoTNos $N$. Observe that the MSE performance degrades upon increasing the dimensionality of the parameter. When increasing the number of IoTNos, the MSE performance improves as it leads to the availability of more measurements of
the parameter, thereby increasing the estimation accuracy. The performance of the proposed SOMP-based hybrid transceiver design is observed to be very close to that of the fully-digital transceiver design, demonstrating the efficiency of the proposed design. \par
%\begin{figure}
%\centering
%\includegraphics[width=\linewidth]{MSE_vs_NRF.eps}
%\label{fig:5}
%\caption{MSE versus number of RF chains ($N_{\text{RF}}^n$) for $N=20$ sensors}
%\end{figure}
%\begin{figure}
%\centering
%\includegraphics[width=\linewidth]{MSE_vs_SNR_heterogeneous.eps}
%\label{fig:6}
%\caption{MSE versus $\text{SNR}_{\text{FC}}$ for different number of receive antennas $N_r \in \left\lbrace10,20\right\rbrace$}
%\end{figure}
\begin{figure*}
\centering
\subfloat[]{\includegraphics[width=0.45\linewidth]{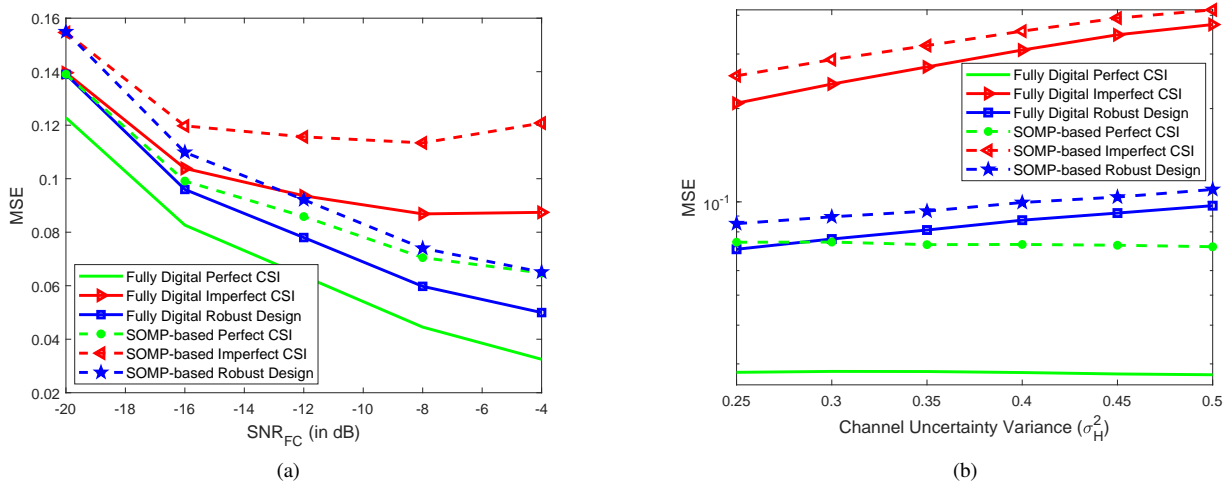}}
\hfil
\hspace{10pt}\subfloat[]{\includegraphics[width=0.45\linewidth]{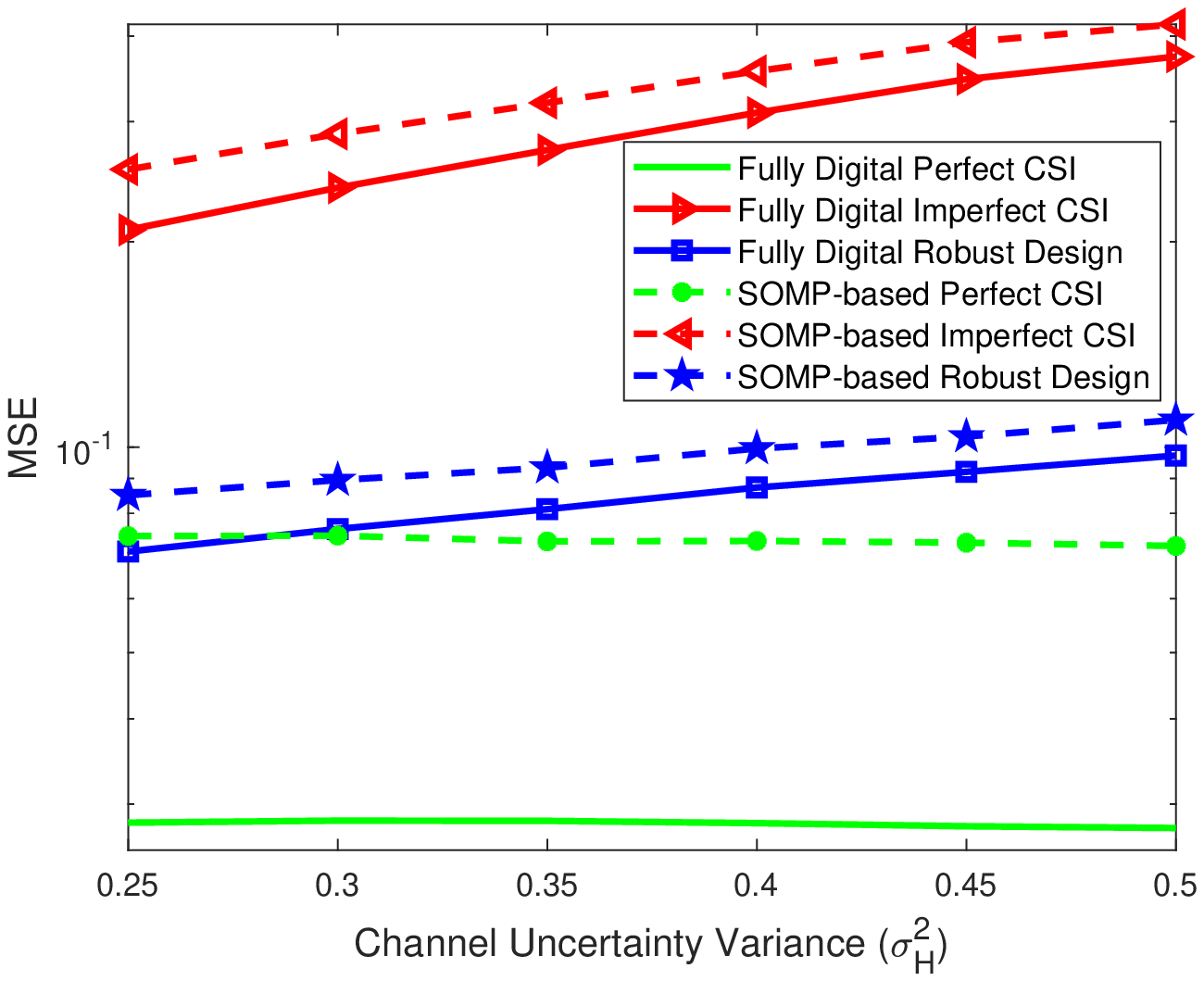}}
\caption{$\left(a\right) $ MSE versus $\text{SNR}_{\text{FC}}$  $\left(b\right)$ MSE versus the channel uncertainty variance $(\sigma_H^2)$ for $N=20$ IoTNos.}
\label{fig:6}
\end{figure*} 
\begin{figure*}
\centering
\subfloat[]{\includegraphics[width=0.45\linewidth]{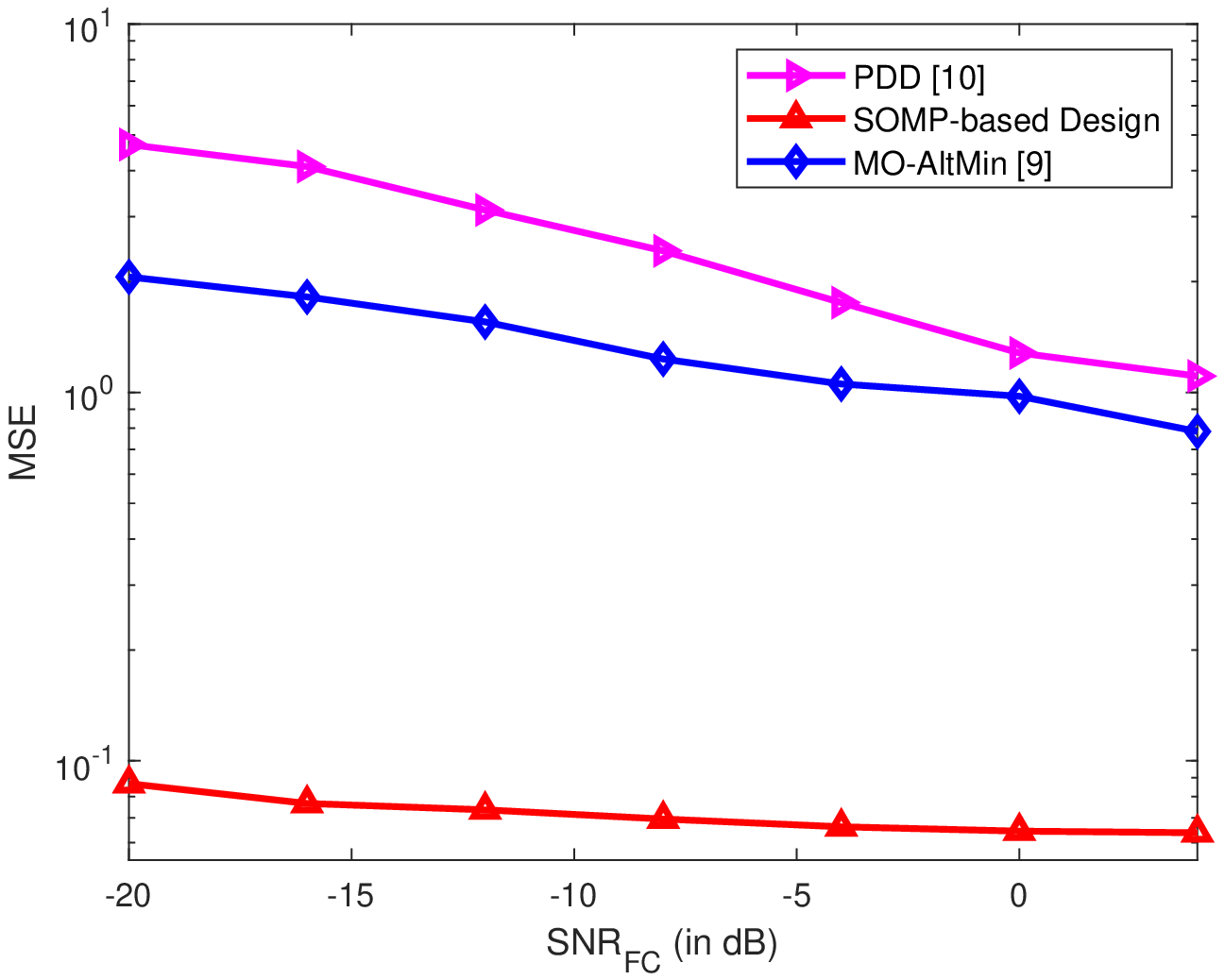}}
\hfil
\hspace{10pt}\subfloat[]{\includegraphics[width=0.45\linewidth]{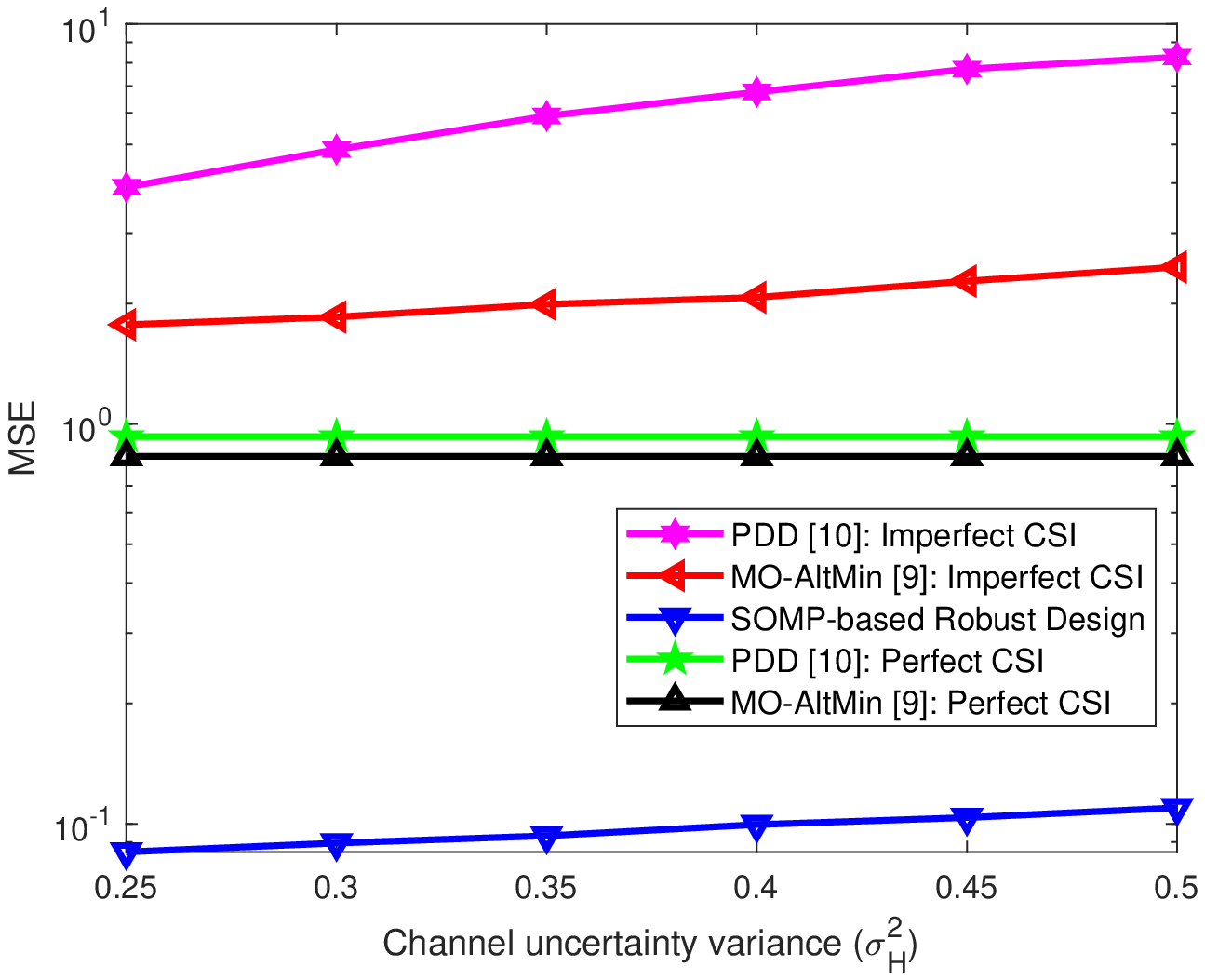}}
\caption{$\left(a\right) $ MSE versus $\text{SNR}_{\text{FC}}$  $\left(b\right)$ MSE versus the channel uncertainty variance $(\sigma_H^2)$ for $N=20$ IoTNos.}
\label{fig:8}
\end{figure*} 
\begin{figure}
\centering
\includegraphics[width=0.95\linewidth]{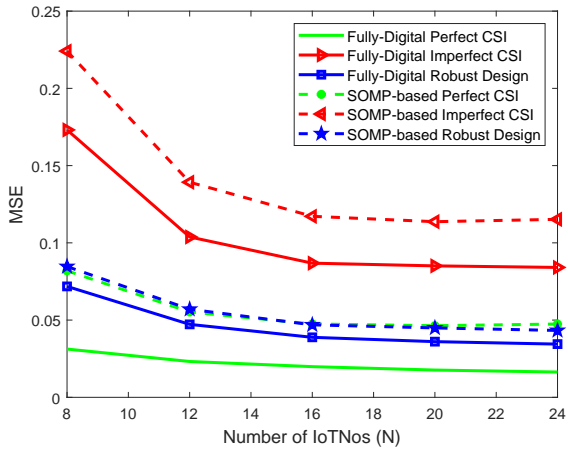}
\label{fig:7}
\caption{MSE versus number of IoTNos ($N$)}
\end{figure}  
Fig. \ref{fig:5}(a) illustrates the MSE versus the number of RF chains $N_{\text{RF}}^n$ at each IoTNo for different number of RF chains $N_{\text{RF}}$ at the noisy FC. The number of IoTNos considered in the system is $N=20$. Again, the MSE performance of the SOMP-based hybrid transceiver improves upon increasing the number of RF chains. It becomes equivalent to the MSE corresponding to the fully-digital design, because $N_{\text{RF}}^n$ becomes equal to the total number of TAs $N_t=5$. It is also observed that upon increasing the number of RF chains at the FC, the MSE performance further improves. However, for $N_{\text{RF}} \geq K$ clusters, it is observed that the MSE performance of the hybrid designs becomes similar. This demonstrates that the FC is indeed capable of recovering the active beams using our SOMP-based RF RC when $N_{\text{RF}} = \text{rank}\left(\mathbf{H}\right)$ and there is no further improvement in performance upon increasing $N_{\text{RF}}$.\par
Fig. \ref{fig:5}(b) illustrates the MSE performance of the hybrid transceiver designs versus $\text{SNR}_{\text{FC}}$. For comparison, a heterogeneous IoTNe is considered, where the observation noise power at each sensor takes values from the following set $\left\lbrace-10,-9,\cdots,9\right\rbrace \text{dB}$. This setup is compared to a homogeneous IoTNe, where the observation noise at each sensor is $\sigma_l^2=-10 \text{dB}$. The homogeneous mmWave IoTNe is observed to perform better than its counterpart if the SNR value at each IoTNo for the homogeneous mmWave IoTNe is higher than that of heterogeneous mmWave counterpart. Upon increasing the number of RAs $N_r$ from $10$ to $20$, the MSE performance improves due to the array gain in the noise limited region. As $\text{SNR}_{\text{FC}}$ increases, the performance gain due to the increasing array gain becomes negligible. \par
Fig. \ref{fig:6}(a) plots the MSE as a function of $\text{SNR}_{\text{FC}}$ for the hybrid TPC/ RC designs derived in Section-\ref{Sec-IV} and Section-\ref{Sec-V}, respectively. The iterative algorithm is run for 15 iterations for each $\text{SNR}_{\text{FC}}$ value. As expected, the MSE performance improves as $\text{SNR}_{\text{FC}}$ increases, and the robust design offers significant performance improvements over the imperfect CSI-based design that ignores the CSI uncertainty. The corresponding performance of the fully-digital transceiver is once again plotted for benchmarking the hybrid transceiver designs.\par 
Fig. \ref{fig:6}(b) shows the MSE performance of the hybrid transceiver designs proposed in Section-\ref{Sec-IV} and Section-\ref{Sec-V}, for scenarios of perfect and imperfect CSI, respectively. To benchmark the performance of the hybrid TPC/ RC designs, the MSE performance corresponding to their fully-digital counterparts is also plotted. The iterative algorithm proposed for our robust transceiver design converges in 15 iterations. It can observed from the figure that the proposed robust design offers an improved MSE performance in comparison to its uncertainty-agnostic counterpart. In addition, its performance is also close to the perfect CSI-based design, which illustrates its value for practical systems. \par
Fig. \ref{fig:8}(a)  compares the MSE performance of our hybrid
TPC/ RC design of Section \ref{Sec-IV} to that of the corresponding designs in \cite{7397861} and \cite{8332507} that proposed the alternating minimization (AltMin) algorithm based on manifold
optimization (MO) and penalty dual decomposition (PDD), respectively. The
figure clearly demonstrates that the proposed SOMP-based
design outperforms both these algorithms. Similarly, Fig. \ref{fig:8}(b) compares the MSE performance of the proposed SOMP-based robust hybrid TPC/ RC design of Section \ref{Sec-V} with that of \cite{7397861} and \cite{8332507}, considering both 
perfect CSI and CSI-uncertainty agnostic approaches. It can be deduced from the figure that the proposed robust design offers a significant MSE performance improvement over both the agnostic as well as perfect CSI-based hybrid designs of \cite{7397861} and \cite{8332507}. This illustrates the efficacy of our proposed robust design.\par
Fig. 8 shows the MSE performance versus the number of IoTNos $N$ in the network for the schemes proposed in Section-IV and -V, both with perfect and imperfect CSI, respectively. As expected, the MSE decreases upon increasing the number of sensors, which reinforces the trend seen in the previous figures.\par

\section{Conclusion} \label{Sec-VII}
Hybrid transceiver designs were conceived for the LDE of a parameter vector in a mmWave MIMO IoTNe. Separate design procedures were presented for a noiseless as well as noisy FC. An important observation for the proposed hybrid transceiver design for a noiseless FC is that even in systems having a large number of antennas both at the IoTNos and the FC, increasing the number of RF chains beyond the parameter dimension does not result in any further MSE improvement. For a noisy FC, the fully digital TPC/ RC are initially determined using a BCD-based iterative framework followed by the SOMP technique for decomposing it into its RF and BB components. A robust hybrid design procedure was also formulated for a practical scenario having CSI uncertainty, which provides improved estimation performance over its CSI uncertainty-agnostic counterpart. Our simulation results illustrate the performance of the proposed schemes and  corroborate our analytical findings.
\begin{appendices}
\begin{figure*}
\begin{align}
\label{MSE_robust_expected}
\mathbb{E}_{\left\lbrace\Delta\mathbf{H}_n\right\rbrace}\zeta\left(\mathbf{A}, \mathbf{P}_{n}\right)&=\bar{\zeta}\left(\mathbf{A},\mathbf{P}_n\right) =\text{Tr}\Bigg[\mathbf{A}\left(\sum_{n=1}^{N}\mathbb{E}_{\Delta\mathbf{H}_n}\left\lbrace\mathbf{H}_n\mathbf{P}_n\left(\mathbf{C}_n\mathbf{R}_{\theta}\mathbf{C}_n^H+\mathbf{R}_n\right)\mathbf{P}^H_n\mathbf{H}^H_n\right\rbrace\right)\mathbf{A}^H+\mathbf{A}\mathbf{R}_n\mathbf{A}^H+\mathbf{R}_{\theta}\nonumber\\ 
&+\mathbf{A}\left(\sum_{n=1}^{N}\sum_{j=1,j\neq n}^{N}\mathbb{E}_{\left\lbrace\Delta\mathbf{H}_n,\Delta\mathbf{H}_j\right\rbrace}\mathbf{H}_n\mathbf{P}_n\mathbf{C}_n\mathbf{R}_{\theta}\mathbf{C}^H_j\mathbf{P}^H_j\mathbf{H}^H_j\right)\mathbf{A}^H-\mathbf{A}\left(\sum_{n=1}^N \mathbb{E}_{\Delta\mathbf{H}_n}\mathbf{H}_n\mathbf{P}_n \mathbf{C}_n\right)\mathbf{R}_{\theta}  \nonumber \\
&-\mathbf{R}_{\theta}\left(\sum_{n=1}^N\mathbb{E}_{\Delta\mathbf{H}_n}\mathbf{C}^H_n\mathbf{P}^H_n\mathbf{H}^H_n\right)\mathbf{A}^H\Bigg]. \nonumber \\
\end{align}
\hrulefill
\end{figure*}
\section{Calculation of dual variables}
\label{appdx:dualv}
To obtain the dual variable $\lambda_n$ at the $n$th IoTNo, let the following matrices be defined as
\begin{align}
&\mathbf{X}_n = \mathbf{H}^H_n\mathbf{A}^H\mathbf{A}\mathbf{H}_n, \\
&\mathbf{Y}_n = \Bigg[\mathbf{H}^H_n\mathbf{A}^H\mathbf{R}_{\theta}\mathbf{G}^H_n-\sum_{j=1,j\neq n}^N\mathbf{H}^H_n\mathbf{A}^H\mathbf{A}\mathbf{H}_j\mathbf{P}_j\nonumber \\
&\left(\mathbf{C}_n\mathbf{R}_{\theta}\mathbf{C}_j^H\right)^H\Bigg]\left[\mathbf{C}_n\mathbf{R}_{\theta}\mathbf{C}^H_n+\mathbf{R}_n\right]^{-1},
\end{align}
where $\mathbf{X}_n \in \mathbb{C} ^{N_t \times N_t}$ and $\mathbf{Y}_n \in \mathbb{C} ^{N_t \times l}$. The minimum-MSE TPC matrix in \eqref{opt_pre} can be compactly written as
\begin{align}
\mathbf{P}^*_n= \left[\mathbf{X}_n+\lambda_n\mathbf{I}\right]^{-1}\mathbf{Y}_n,
\end{align}
Since $\mathbf{X}_n$ is a Hermitian symmetric matrix, its eigenvalue decomposition (EVD) is given by $\widetilde{\mathbf{U}}_n\widetilde{\mathbf{\Lambda}}_n\widetilde{\mathbf{U}}_n^H$. Initially, $\mathbf{P}^*_n$ is calculated by setting $\lambda_n = 0$. If the average transmit power in \eqref{Pow_Cons} is satisfied, then $\lambda_n = 0$. Otherwise for a non-zero dual variable, the average transmit power is met with equality from the complementary slackness condition
\begin{align}
\lambda_n\left[\text{Tr}\left[\mathbf{P}_{n}\left(\mathbf{C}_n\mathbf{R}_{\boldsymbol{\theta}}\mathbf{C}^H_n+\mathbf{R}_n\right)\mathbf{P}^H_{n}\right]- \rho_n\right]=0.
\end{align}
The value of the dual variable $\lambda_n$ for the average transmit power to be satisfied with equality can be calculated as
\begin{align}
&\text{Tr}\left[\mathbf{P}_{n}\left(\mathbf{C}_n\mathbf{R}_{\boldsymbol{\theta}}\mathbf{C}^H_n+\mathbf{R}_n\right)\mathbf{P}^H_{n}\right]=\rho_n \\ \nonumber
&=\text{Tr}\left[\bigg[\mathbf{X}_n+\lambda_n\mathbf{I}\right]^{-1}\mathbf{Y}_n\left(\mathbf{C}_n\mathbf{R}_{\boldsymbol{\theta}}\mathbf{C}^H_n+\mathbf{R}_n\right) \\ \nonumber
&\ \ \times \mathbf{Y}_n^H\left[\mathbf{X}_n+\lambda_n\mathbf{I}\right]^{-1}\bigg]\\ \nonumber
&=\text{Tr}\bigg[\left[\widetilde{\mathbf{U}}_n\widetilde{\mathbf{\Lambda}}_n\widetilde{\mathbf{U}}_n^H+\lambda_n\mathbf{I}\right]^{-1}\mathbf{Y}_n\left(\mathbf{C}_n\mathbf{R}_{\boldsymbol{\theta}}\mathbf{C}^H_n+\mathbf{R}_n\right)\\ \nonumber
&\ \  \times \mathbf{Y}_n^H\left[\widetilde{\mathbf{U}}_n\widetilde{\mathbf{\Lambda}}_n\widetilde{\mathbf{U}}_n^H+\lambda_n\mathbf{I}\right]^{-1}\bigg]\\ \nonumber
&=\text{Tr}\bigg[\widetilde{\mathbf{U}}_n\left[\widetilde{\mathbf{\Lambda}}_n+\lambda_n\mathbf{I}\right]^{-1}\widetilde{\mathbf{U}}_n^H\mathbf{Y}_n\left(\mathbf{C}_n\mathbf{R}_{\boldsymbol{\theta}}\mathbf{C}^H_n+\mathbf{R}_n\right)\\ \nonumber
&\ \  \times \mathbf{Y}_n^H\widetilde{\mathbf{U}}_n\left[\mathbf{\Lambda}_\text{n}+\lambda_n\mathbf{I}\right]^{-1}\widetilde{\mathbf{U}}_n^H\bigg]\\ \nonumber
&=\text{Tr}\bigg[\left[\widetilde{\mathbf{\Lambda}}_n+\lambda_n\mathbf{I}\right]^{-2} \times\\ \nonumber
&\ \ \ \ \underbrace{\widetilde{\mathbf{U}}_n^H\mathbf{Y}_n\left(\mathbf{C}_n\mathbf{R}_{\boldsymbol{\theta}}\mathbf{C}^H_n+\mathbf{R}_n\right)\mathbf{Y}_n^H\widetilde{\mathbf{U}}_n}_{\mathbf{Z}_n} \bigg] \\ \nonumber
&\ \ \ \ = \sum_{k=1}^{N_t} \frac{z_{kk}}{\left(\tilde{\lambda}_{k}+\lambda_n\right)^2}=\rho_n,
\end{align}
where $\tilde{\lambda}_{k}$ is the $k$th eigenvalue of $\mathbf{X}_n$ and $z_{kk}$ is the $\left(k,k\right)$th element of matrix $\mathbf{Z}_n= \widetilde{\mathbf{U}}_n^H\mathbf{Y}_n\left(\mathbf{C}_n\mathbf{R}_{\boldsymbol{\theta}}\mathbf{C}^H_n+\mathbf{R}_n\right)\mathbf{Y}_n^H\widetilde{\mathbf{U}}_{n}$. The dual variable $\lambda_n$ can now be obtained via the classic bisection method.
\begin{figure*}
\begin{align} 
\bar{\zeta}\left(\mathbf{A},\mathbf{P}_n\right)&=\text{Tr}\Bigg[\mathbf{A}\bigg(\sum_{n=1}^{N}\widehat{\mathbf{H}}_n\mathbf{P}_n\left(\mathbf{C}_n\mathbf{R}_{\theta}\mathbf{C}_n^H+\mathbf{R}_n\right)\mathbf{P}^H_n\widehat{\mathbf{H}}^H_n\bigg)\mathbf{A}^H +\bigg(\sum_{n=1}^N\sigma_H^2\text{Tr}\left(\mathbf{C}_n\mathbf{R}_{\theta}\mathbf{C}_n^H+\mathbf{R}_n\right)\bigg)\mathbf{A}\mathbf{A}^H+ \mathbf{A}\mathbf{R}_n\mathbf{A}^H+\nonumber\\ 
&\mathbf{R}_{\theta}+\mathbf{A}\left(\sum_{n=1}^{N}\sum_{j=1,j\neq n}^{N}\widehat{\mathbf{H}}_n\mathbf{P}_n\mathbf{C}_n\mathbf{R}_{\theta}\mathbf{C}^H_j\mathbf{P}^H_j\widehat{\mathbf{H}}^H_j\right)\mathbf{A}^H \nonumber -\mathbf{A}\left(\sum_{n=1}^N \widehat{\mathbf{H}}_n\mathbf{P}_n \mathbf{C}_n\right)\mathbf{R}_{\theta}-\mathbf{R}_{\theta}\left(\sum_{n=1}^N\widehat{\mathbf{C}}^H_n\mathbf{P}^H_n\mathbf{H}^H_n\right)\mathbf{A}^H \nonumber \Bigg].
\end{align} \label{last}
\hrulefill
\end{figure*}
\section{Average MSE calculation in \eqref{MSE_Robust}}
\label{appdx:MSE_Robust}
The channel estimation error matrices $\left\lbrace\Delta\mathbf{H}_n\right\rbrace_{n=1}^{N}$ for each IoTNo in \eqref{EQ:H} are assumed to be independent of each other. The average MSE $\bar{\zeta}\left(\mathbf{A},\mathbf{P}_n\right)$ is given by \eqref{MSE_robust_expected} shown at the top of this page.

The lemma described below is employed to evaluate the average MSE.
\begin{lem}
\label{Lemma1}
Consider a matrix $\mathbf{H}_n \in \mathbb{C}^{N_R \times N_T}$, defined as $\mathbf{H}_n=\widehat{\mathbf{H}}_n+\Delta{\mathbf{H}}_n$, where $\widehat{\mathbf{H}}_n$ is known and each element of $\Delta{\mathbf{H}}_n$ is assumed to be Gaussian distributed as $\mathcal{CN}\left(0,\sigma_H^2\right)$. Then, for a matrix $\mathbf{K}$, the following properties hold
\begin{align*}
&\mathbb{E}_{\Delta\mathbf{H}_n}\left\lbrace\mathbf{H}_n\mathbf{K}\mathbf{K}^H\mathbf{H}_n^H\right\rbrace = \widehat{\mathbf{H}}_n\mathbf{K}\mathbf{K}^H\widehat{\mathbf{H}}_n^H + \sigma_H^2\mathrm{Tr}\left(\mathbf{K}\mathbf{K}^H\right)\mathbf{I}_{N_R} \\
&\mathbb{E}_{\Delta\mathbf{H}_n,\Delta\mathbf{H}_j}\left\lbrace\mathbf{H}_n\mathbf{K}\mathbf{K}^H\mathbf{H}_j^H\right\rbrace = \widehat{\mathbf{H}}_n\mathbf{K}\mathbf{K}^H\widehat{\mathbf{H}}_j^H, n \neq j. \\
\end{align*}
\end{lem}
\begin{proof}
The first result of Lemma \ref{Lemma1} can be proved as follows:
The term $\mathbf{H}_n\mathbf{K}\mathbf{K}^H\mathbf{H}_n^H$ can be expanded as
\begin{align}
\label{Channel_robust}
&\mathbf{H}_n\mathbf{K}\mathbf{K}^H\mathbf{H}_n^H = \left(\widehat{\mathbf{H}}_n+\Delta\mathbf{H}_n\right)\mathbf{K}\mathbf{K}^H\left(\widehat{\mathbf{H}}_n+\Delta\mathbf{H}_n\right)^H \nonumber \\ 
&= \widehat{\mathbf{H}}_n\mathbf{K}\mathbf{K}^H\widehat{\mathbf{H}}_n^H + \underbrace{{\Delta\mathbf{H}}_n\mathbf{K}\mathbf{K}^H\widehat{\mathbf{H}}_n^H}_{\mathbf{T}_1} +\underbrace{\widehat{\mathbf{H}}_n\mathbf{K}\mathbf{K}^H{\Delta\mathbf{H}}_n^H}_{\mathbf{T}_2} \nonumber \\
&\ \ +\underbrace{{\Delta\mathbf{H}}_n\mathbf{K}\mathbf{K}^H{\Delta\mathbf{H}}_n^H}_{\mathbf{T}_3}.
\end{align}
For compact representation, consider $\mathbf{T}_1,\mathbf{T}_2$ and $\mathbf{T}_3$ as shown above. \newline
Each $\left(a,b\right)\text{th}$ element of $\mathbf{T}_1$ is the dot product of the $a\text{th}$ row of ${\Delta\mathbf{H}}_n$ and $b\text{th}$ column of $\mathbf{K}\mathbf{K}^H\widehat{\mathbf{H}}_n^H$. This implies that
\begin{align*}
\left[\mathbf{T}_1\right]_{ab}=\sum_{i=1}^{N_T}\left[{\Delta\mathbf{H}}_n\right]_{ai}\left[{\mathbf{K}}\mathbf{K}^H\widehat{\mathbf{H}}_n^H\right]_{ib}.
\end{align*} 
Thus, the expected value of any element of $\mathbf{T}_1$ is zero, since $\mathbb{E}\left[{\Delta\mathbf{H}}_n\right]_{ai}=0, 1\leq a \leq N_R, 1\leq i \leq N_T$. This further implies that
\begin{align*}
\mathbb{E}_{\Delta\mathbf{H}_n}\left\lbrace\mathbf{T}_1\right\rbrace=\mathbf{0}.
\end{align*}
\newline
Following the same reasoning as above, it can be shown that
\begin{align*}
\mathbb{E}_{\Delta\mathbf{H}_n}\left\lbrace\mathbf{T}_2\right\rbrace=\mathbf{0}.
\end{align*}
\newline
To compute the expected value of $\mathbf{T}_3$, let us denote  the EVD of the Hermitian matrix $\mathbf{K}\mathbf{K}^H$ as $\mathbf{U\Lambda}\mathbf{U}^H$. Therefore, we have
\begin{align*}
\mathbb{E}_{\Delta\mathbf{H}_n}\left\lbrace\mathbf{T}_3\right\rbrace&=\mathbb{E}_{\Delta\mathbf{H}_n}\left\lbrace\Delta\mathbf{H}_n\mathbf{U\Lambda}\mathbf{U}^H{\Delta\mathbf{H}}_n^H\right\rbrace \\
&=\mathbb{E}_{\Delta\widetilde{\mathbf{H}}_n}\left\lbrace\Delta\widetilde{\mathbf{H}}_n\mathbf{\Lambda}{\Delta\widetilde{\mathbf{H}}}_n^H\right\rbrace, 
\end{align*}
where $\Delta\widetilde{\mathbf{H}}_n=\Delta\mathbf{H}_n\mathbf{U}$, whose elements are also distributed as $\mathcal{CN}\left(0,\sigma_H^2\right)$ since $\mathbf{U}$ is a unitary matrix. Thus, the above expression can be simplified as follows:
\begin{align*}
\mathbb{E}_{\Delta\widetilde{\mathbf{H}}_n}\left\lbrace\Delta\widetilde{\mathbf{H}}_n\mathbf{\Lambda}{\Delta\widetilde{\mathbf{H}}}_n^H\right\rbrace = \mathbb{E}_{\Delta\widetilde{\mathbf{H}}_n}\left\lbrace
\sum_{i=1}^{N_T}\tilde{\lambda}_i\Delta\widetilde{\mathbf{h}}_{n,i}{\Delta\widetilde{\mathbf{h}}}_{n,i}^H\right\rbrace,
\end{align*}
where $\Delta\widetilde{\mathbf{h}}_{n,i}$ denotes the $i\text{th}$ column of $\Delta\widetilde{\mathbf{H}}_n$. Upon assuming that all the elements of $\Delta\mathbf{H}_n$ are independent and identically distributed, $\mathbb{E}_{\Delta\widetilde{\mathbf{H}}_n}\left\lbrace
\Delta\widetilde{\mathbf{h}}_{n,i}{\Delta\widetilde{\mathbf{h}}}_{n,i}^H\right\rbrace = \sigma_H^2\mathbf{I}_{N_R}$, we have
\begin{align*}
\mathbb{E}_{\Delta\widetilde{\mathbf{H}}_n}\left\lbrace
\sum_{i=1}^{N_T}\tilde{\lambda}_i\Delta\widetilde{\mathbf{h}}_{n,i}{\Delta\widetilde{\mathbf{h}}}_{n,i}^H\right\rbrace &= \sum_{i=1}^{N_T}\tilde{\lambda}_i\mathbb{E}_{\Delta\widetilde{\mathbf{H}}_n}\left\lbrace
\Delta\widetilde{\mathbf{h}}_{n,i}{\Delta\widetilde{\mathbf{h}}}_{n,i}^H\right\rbrace \\
&= \sum_{i=1}^{N_T}\tilde{\lambda}_i\sigma_H^2\mathbf{I}_{N_R} \\
&= \sigma_H^2\text{Tr}\left(\mathbf{K}\mathbf{K}^H\right)\mathbf{I}_{N_R}.
\end{align*}
From the above results and \eqref{Channel_robust}, we conclude that
\begin{align*}
&\mathbb{E}_{\Delta\mathbf{H}_n}\left\lbrace\mathbf{H}_n\mathbf{K}\mathbf{K}^H\mathbf{H}_n^H\right\rbrace = \mathbb{E}_{\Delta\mathbf{H}_n}\left\lbrace\widehat{\mathbf{H}}_n\mathbf{K}\mathbf{K}^H\widehat{\mathbf{H}}_n^H\right\rbrace +  \\
&\ \  \mathbb{E}_{\Delta\mathbf{H}_n}\left\lbrace\mathbf{T}_1\right\rbrace+\mathbb{E}_{\Delta\mathbf{H}_n}\left\lbrace\mathbf{T}_2\right\rbrace+ \mathbb{E}_{\Delta\mathbf{H}_n}\left\lbrace\mathbf{T}_3\right\rbrace\\
&  \ \ =  \widehat{\mathbf{H}}_n\mathbf{K}\mathbf{K}^H\widehat{\mathbf{H}}_n^H + \mathbf{0}+ \mathbf{0}+ \sigma_H^2\text{Tr}\left(\mathbf{K}\mathbf{K}^H\right)\mathbf{I}_{N_R}\\
&  \ \ =  \widehat{\mathbf{H}}_n\mathbf{K}\mathbf{K}^H\widehat{\mathbf{H}}_n^H + \sigma_H^2\text{Tr}\left(\mathbf{K}\mathbf{K}^H\right)\mathbf{I}_{N_R}.
\end{align*}
Hence, the first result of Lemma \ref{Lemma1} is proved.
\par
The next result of the lemma can be proven by expanding the quantity $\mathbf{H}_n\mathbf{K}\mathbf{K}^H\mathbf{H}_j^H$ as
\begin{align}
\label{Channel_robust_1}
\mathbf{H}_n\mathbf{K}\mathbf{K}^H\mathbf{H}_j^H &= \widehat{\mathbf{H}}_n\mathbf{K}\mathbf{K}^H\widehat{\mathbf{H}}_j^H + {\Delta\mathbf{H}}_n\mathbf{K}\mathbf{K}^H\widehat{\mathbf{H}}_j^H  \nonumber \nonumber \\
&\ \ +\widehat{\mathbf{H}}_n\mathbf{K}\mathbf{K}^H{\Delta\mathbf{H}}_j^H+{\Delta\mathbf{H}}_n\mathbf{K}\mathbf{K}^H{\Delta\mathbf{H}}_j^H.
\end{align}
As explained previously, the expected value of the second and third terms above is equal to zero. The expected value of the fourth term is given by
\begin{align*}
&\mathbb{E}_{\Delta\mathbf{H}_n,\Delta\mathbf{H}_j}\left\lbrace\Delta\mathbf{H_n}\mathbf{K}\mathbf{K}^H\Delta\mathbf{H}_j^H\right\rbrace \\
&\ \  \stackrel{a}{=}\mathbb{E}_{\Delta\mathbf{H}_n}\left\lbrace\Delta\mathbf{H_n}\right\rbrace\mathbf{K}\mathbf{K}^H\mathbb{E}_{\Delta\mathbf{H}_j}\left\lbrace\Delta\mathbf{H}_j^H\right\rbrace \\
&\ \ =\mathbf{0},
\end{align*}
where (a) follows from the fact that $\Delta\mathbf{H}_n$ and $\Delta\mathbf{H}_j$ are independent of each other. This implies that
\begin{align*}
\mathbb{E}_{\Delta\mathbf{H}_n,\Delta\mathbf{H}_j}\left\lbrace\mathbf{H}_n\mathbf{K}\mathbf{K}^H\mathbf{H}_j^H\right\rbrace =  \widehat{\mathbf{H}}_n\mathbf{K}\mathbf{K}^H\widehat{\mathbf{H}}_j^H,
\end{align*}
which proves the second result of Lemma \ref{Lemma1}. 
\end{proof}
Now, upon employing Lemma \ref{Lemma1} in \eqref{MSE_robust_expected}, one obtains the desired expression given at the top of this page.
\end{appendices}
\bibliographystyle{IEEEtran}
\bibliography{manuscript}
\end{document}